\documentclass[11pt]{article}

\usepackage[utf8]{inputenc}
\usepackage[T1]{fontenc}
\usepackage[margin=1in]{geometry}
\usepackage{graphicx}
\usepackage{amsmath,amssymb}
\usepackage{natbib}
\usepackage{subcaption}
\usepackage{xcolor}
\usepackage[colorlinks=true,linkcolor=blue,citecolor=blue,urlcolor=blue]{hyperref}
\usepackage{authblk}

\title{A Fast and Physically Grounded Ocean Model for GCMs: The Dynamical Slab Ocean Model of the Generic-PCM (rev. 3423)}

\author[1,2,3]{Siddharth Bhatnagar}
\author[7]{Francis Codron}
\author[4]{Ehouarn Millour}
\author[1,2]{Emeline Bolmont}
\author[2,3]{Maura Brunetti}
\author[2,3]{Jérôme Kasparian}
\author[4,5]{Martin Turbet}
\author[2,6]{Guillaume Chaverot}

\affil[1]{Observatoire astronomique de l’Université de Genève, Versoix, Switzerland}
\affil[2]{Centre pour la vie dans l'Univers de l'Université de Genève, Genève, Switzerland}
\affil[3]{Group of Applied Physics and Institute for Environmental Sciences, University of Geneva, Geneva, Switzerland}
\affil[4]{Laboratoire de Météorologie Dynamique/IPSL, CNRS, Sorbonne Université, École Normale Supérieure, PSL Research University, École Polytechnique, 75005 Paris, France}
\affil[5]{Laboratoire d'astrophysique de Bordeaux, Univ. Bordeaux, CNRS, B18N, allée Geoffroy Saint-Hilaire, 33615 Pessac, France}
\affil[6]{Univ. Grenoble Alpes, CNRS, IPAG, F-38000 Grenoble, France}
\affil[7]{Laboratoire d'Océanographie et du Climat LOCEAN/IPSL, Sorbonne Université, CNRS, MNHN, IRD, 75005 Paris, France}

\date{}

\begin{document}
\maketitle

\begin{abstract}
We present an improved dynamical slab ocean model implemented in a 3-D General Circulation Model (GCM) called the Generic Planetary Climate Model (Generic-PCM; formerly the LMD-Generic GCM). 
Earlier two-layer slab ocean models featured wind-driven Ekman transport, horizontal diffusion and convective adjustment.
Building upon this, our updated parallelised model introduces a Sverdrup balance scheme for Ekman transport, the first application of the Gent–McWilliams (GM) parameterisation of mesoscale eddies in a slab ocean model, and a spectrally and thickness dependent formulation of sea ice and snow albedo.
We validate this model in an idealised aquaplanet setting under various OHT configurations. We show that enabling OHT transforms not only surface features -- such as cooler tropical sea surface temperatures (SSTs) and reduced sea ice coverage -- but also atmospheric structures, notably producing a double-banded precipitation pattern across the equator driven by Ekman-induced upwelling. Our modelled meridional OHT profiles show first-order agreement with fully coupled atmosphere-ocean GCMs, with Ekman transport dominating in the tropics (enhanced by GM-induced restratification), and diffusive plus GM contributions peaking near the ice edge.
When applied to modern Earth, the OHT-enabled configuration yields an annual global average surface temperature of 13°C, within 1°C of reanalysis estimates, and improves extrapolar SSTs and sea ice coverage relative to the OHT-disabled baseline. Seasonal SST and sea ice biases relative to observations are also significantly reduced to within 0.6°C and 3 million km², respectively. We obtain a planetary bond albedo of around 0.32, in close agreement with observations. 
Together, the aquaplanet and modern Earth benchmarks demonstrate that our developments represent a clear improvement over earlier two-layer implementations.
We further show that GM-induced restratification reduces the need for explicit convective adjustment, while also strengthening Ekman transport. In addition to improving equatorial dynamics, the inclusion of the Sverdrup balance also reduces hemispheric asymmetries.
Notably, due to model parallelisation, these improvements are achieved at almost no additional computational cost compared to OHT-disabled simulations run over the same number of model years. This enables long integrations and large ensemble studies, making the model particularly well suited for exoplanet and paleoclimate studies where broad parameter exploration is essential.
\end{abstract}

\textit{Accepted for publication in Geoscientific Model Development (GMD)}

\section{Introduction}



Oceans are central to climate regulation on terrestrial planets due to their ability to store and redistribute heat. On Earth, ocean heat transport (OHT) plays a particularly crucial role: it carries excess thermal energy from the equator toward the poles, significantly reducing meridional temperature gradients. Observational estimates clearly reveal this latitudinal energy redistribution \citep[e.g.,][]{trenberth1994global}, which is accomplished through a combination of shallow wind-driven gyres in the tropics and subtropics, and deeper overturning circulations linked to high-latitude convection \citep[e.g.,][]{trenberth2001estimates,marshall_plumb_2007}. 
Together with the ocean’s high thermal inertia and interactions with sea ice, these processes help dampen seasonal and regional temperature extremes — stabilising Earth’s climate over long timescales and contributing fundamentally to its long-term habitability.


Ocean dynamics are also critical in other planetary contexts -- including paleoclimates and exoplanets. Paleoclimate simulations have shown that OHT strongly influences transitions between climate states. For example, OHT can effectively delay or prevent global glaciation by limiting sea ice expansion in Snowball Earth scenarios \citep[for e.g.,][]{pierrehumbert2011climate,yang2012climatedynamics,yang2012solarco2}. Dynamic oceans have also been incorporated in studies of ancient climates on Earth \citep[e.g.,][]{olson2022effect,ragon2024alternative}, Venus \citep[e.g.,][]{way2016venus,way2020venusian} and Mars \citep[e.g.,][]{schmidt2022circumpolar}, where they play a key role in determining planetary climate regimes and potential habitability. \citet{rose2015} and \citet{brunetti2019co} demonstrated that the choice of oceanic representation can even determine the number and nature of climatic steady states.



Most known habitable zone \citep[HZ;][]{kasting1993habitable} exoplanets orbit low-mass stars and are expected to undergo tidal spin-down, often settling into a synchronously rotating (tidally locked in a 1:1 spin-orbit resonance; for e.g., see \citealt{goldreich1966spin} and \citealt{pierrehumbert2019atmospheric}), having a permanent dayside and nightside \citep{ribas2016habitability,turbet2016habitability,turbet2018modeling,turbet2020review,braam2025earth} like Proxima-b\footnote{However, the spin state of Proxima-b remains uncertain with studies suggesting a 3:2 spin-orbit resonance capture even for modest orbital eccentricities \citep[e.g.,][]{goldreich1966spin,dobrovolskis2007spin,ribas2016habitability,pierrehumbert2019atmospheric,valente2022tidal}.} \citep{anglada2016terrestrial} and TRAPPIST-1e \citep{gillon2017seven}. Numerous General Circulation Model (GCM) studies have investigated their atmospheric dynamics \citep[for e.g.,][]{edson2011atmospheric,pierrehumbert2010palette,leconte2013increased,kopparapu2016inner,turbet2016habitability,kopparapu2017habitable,boutle2017exploring,taniguchi2026atmospheric}, but few have included dynamic oceans. Yet, the few studies that do incorporate OHT on synchronously rotating planets have demonstrated its profound impact on the climate -- including broadening the area of surface liquid water \citep{hu2014role} and modifying the large-scale circulation, surface temperatures and sea ice patterns \citep{del2019habitable}. \citet{cullum2016importance} highlighted the role of salinity in shaping ocean circulation and the climate at large. \citet{checlair2019no} showed that due to OHT, habitable synchronously rotating planets are unlikely to have snowball states.
The implications of OHT of a few other studies \citep[e.g.,][]{yang2020transition,olson2020oceanographic,batra2024climatic} is further detailed in Sect.~\ref{sec:discussion}. 
All these effects are important not only for assessing planetary habitability but also for interpreting exoplanet observables: for instance, \citet{yang2019ocean} demonstrated that OHT can shift the thermal hotspot eastward in phase curves of temperate synchronously rotating planets. 

These modelling insights are particularly timely, as we enter an era of high-precision exoplanet observations, probing the atmospheres — and soon, the surface conditions — of potentially habitable exoplanets. Small HZ rocky planets are now being discovered and are prime targets for atmospheric characterisation. Ongoing efforts include the James Webb Space Telescope \citep[JWST;][]{gardner2006james}, and near-future ground-based instruments such as RISTRETTO \citep{lovis2024ristretto} planned for the Very Large Telescope (VLT) and the ArmazoNes high Dispersion Echelle Spectrograph \citep[ANDES;][]{palle2025ground}
at the Extremely Large Telescope (ELT). These efforts will be complemented by future space-based observatories like the Habitable Worlds Observatory \citep[HWO;][]{gaudi2020habitable} and the Large Interferometer For Exoplanets \citep[LIFE;][]{quanz2022large}, which aim to detect and characterize the atmospheres of temperate terrestrial exoplanets. Understanding the climate dynamics of such planets and assessing their potential habitability is a major focus of both theoretical and observational research.

Most dynamic ocean studies to date have relied on fully coupled Atmosphere–Ocean General Circulation Models (AOGCMs). These represent the gold standard of physical realism, explicitly solving the Navier-Stokes equations in both the ocean and atmosphere. However, this comes at the cost of computational expense and they require long integration times to reach equilibrium, making them impractical for large parameter space studies. As a compromise, most exoclimate studies opt for lower-level ocean representations described below.

The simplest way to model oceans is as a surface with a higher thermal inertia than land \citep[for e.g.,][]{spiegel2008habitable,dressing2010habitable,bolmont2016habitability}. 
A step up are slab ocean models without heat transport, where the ocean is represented as a single wind-mixed layer with a homogenous temperature that is forced by local surface heat fluxes. This is arguably the most popular ocean modelling technique in exoplanet GCMs \citep[for e.g.,][]{kopparapu2016inner,wolf2017constraints,turbet2021mars,chaverot2023first}. 
A further refinement involves prescribing slab ocean models with meridional heat fluxes (q-fluxes) based on observations / AOGCM simulations \citep[e.g.,][]{wolf2015evolution}. This approach lacks flexibility as the OHT is imposed rather than emergent (see Sect.~\ref{subsec:q-flux}).
Another step up are slab ocean models which simulate OHT purely through a diffusive flux \citep[for e.g.,][]{donnadieu2006modelling,kilic2017multiple}. However, this approach misrepresents the structure of real ocean heat transport, notably failing to capture key features such as the equatorial cold tongue -- the pronounced minimum in equatorial sea surface temperatures \citep{deconto2003rapid}. 

This paper presents a follow-up of the model originally introduced by \citet{codron2012ekman} and later adapted by \citet{charnay2013exploring}. These earlier models included horizontal diffusion and convection, with \citet{codron2012ekman} introducing, for the first time in a slab ocean model, wind-driven Ekman transport -- the dominant driver of tropical meridional OHT in Earth observations \citep{levitus1987meridional,forget2019global}. Building upon this foundation, our work refines the two core processes historically handled by the dynamical slab ocean:
\begin{enumerate}
    \item \textbf{Sea ice evolution}; with a spectral albedo representation, refined treatment of ice formation and melting, and the introduction of a finite heat capacity for snow (Sect.~\ref{subsec:sea-ice-evolution})
    \item \textbf{Heat transport by ocean circulation}; through optimised horizontal diffusion and Ekman transport, a Sverdrup balance scheme near the equator for more consistent tropical meridional transport (Sect.~\ref{subsubsec:ekman-transport}), and the implementation of mesoscale eddy mixing via the Gent-McWilliams parameterisation \citep[][Sect.~\ref{subsubsec:gent-mcwilliams}]{gent1990isopycnal}, marking its first application within a slab ocean model.
\end{enumerate}

Like in \citet{codron2012ekman} and \citet{charnay2013exploring}, our updated model does not include a prognostic salinity field or density-driven overturning circulation. While these processes are important in full AOGCMs \citep[e.g.,][]{cullum2016importance}, they require solving additional tracer and density equations, which substantially increases computational cost. Our model therefore focuses on wind-driven and eddy heat transport processes that capture the first-order meridional OHT while preserving computational efficiency.

The model is fully parallelised, enabling substantially faster integrations than earlier implementations and making it well suited for long simulations and ensemble studies, important in exoplanet and paleoclimate science. With these developments, our dynamical slab ocean model acts as a bridge between higher-level slab-oceans and fully coupled AOGCMs.


In this article, we first present the model in detail, highlighting the improvements made in sea ice evolution and OHT (Sects.~\ref{subsec:sea-ice-evolution} and \ref{subsec:heat-transport} respectively). We then validate it against two benchmark cases: an aquaplanet (Sect.~\ref{sec:aquaplanet}), and modern Earth (Sect.~\ref{sec:earth}). Finally, we discuss the implications of our model, its strengths and limitations in Sect.~\ref{sec:discussion}.

\section{\label{sec:model-description}Model description}

The simulations presented here were conducted with rev. 3423 of the 3-D Generic Planetary Climate Model (Generic-PCM, formerly the LMD-Generic GCM, Forget et al., to be submitted). Originally developed to study present-day Earth’s climate \citep[e.g.,][]{hourdin2006lmdz4,charnay2013exploring} and Mars \citep{forget1999improved}, the model has since been generalised to simulate the (paleo-)climates of other solar system bodies \citep[e.g.,][]{charnay2013exploring,forget20133d,wordsworth2013global,turbet2021day} as well as exoplanets \citep[e.g.,][]{wordsworth2011gliese,bolmont2016habitability,turbet2016habitability,turbet2018modeling,kuzucan2025role}. Notably, it remains the only GCM capable of treating water vapour as a major atmospheric constituent, a crucial feature for investigating water-rich planetary environments \citep[for e.g.,][]{leconte2013increased}. The associated flexibility is exemplified by studies that push the model to extreme scenarios; for instance, in its radiative transfer: simulating climates during \citep{chaverot2023first} and post runaway greenhouse \citep{turbet2021day}, or, its numerical scheme: testing the effect of very large bolide impacts on the climate of early Mars \citep{turbet2020environmental}. 

The 3-D dynamical core of the model \citep{hourdin2006lmdz4} solves the primitive equations of geophysical fluid dynamics in the atmosphere using a finite-difference method on an Arakawa C-grid. The GCM includes a comprehensive radiative transfer scheme accounting for absorption and scattering by atmospheric gases, clouds, and surface interactions, along with surface and subsurface processes such as condensation, evaporation, sublimation, and precipitation.

All simulations in this study use a horizontal resolution of 64$\times$48 grid points (longitude$\times$latitude; 5.625°$\times$3.75°). The atmosphere is discretised into 30 vertical layers extending from the surface up to 10 Pa, using hybrid $\sigma$-coordinates (where $\sigma$ = pressure / surface pressure). Radiative transfer is based on the correlated-k method \citep{fu1992correlated}. We use tables from \cite{leconte2013increased}, computed using the HITRAN 2012 line lists \citep{rothman2013hitran2012} across 38 thermal infrared and 36 visible spectral bands. Water vapour (H$_2$O) is treated as a variable component capable of phase changes from the surface, ocean or atmosphere, with relative humidity computed self-consistently. Sub-grid scale processes, including turbulent mixing and moist convection, follow parameterisations described in \cite{leconte2013increased}. Cloud condensation nuclei (CCN) radii are fixed at 12 $\mu$m for liquid water and 35 $\mu$m for ice, following \cite{charnay2013exploring}. During the spin-up phase, our study keeps time-steps short to accurately capture rapid processes: the dynamical time-step (e.g., for winds) is 45 model seconds, the physical time-step (e.g., for evaporation) is 7.5 min, and the radiative time-step is 45 min. Post spin-up, all time-steps are doubled to improve computational efficiency.

Our dynamical slab ocean model builds on the works of \citet{codron2012ekman} and \citet{charnay2013exploring}. The model solves prognostic equations for ocean temperature and sea ice evolution on the same horizontal grid as the GCM's atmosphere and comprises two ocean layers. The upper layer represents the mixed surface layer interacting with the atmosphere, while the lower layer represents the deeper ocean, exchanging heat with the mixed layer but not directly with the atmosphere -- this two-layer configuration is key for facilitating heat transport. We compute the vertical heat flux using an upstream scheme while a centered scheme is used in the meridional direction. Sea-ice fraction, thickness, and snow thickness are tracked at each ocean grid point. We set the coupling time-step between the ocean and atmosphere (“ocean time-step”) to match the physical time-step (either 7.5 or 15 model minutes, as discussed earlier).

\subsection{\label{subsec:general-principle}General principle}

At the heart of the dynamical slab ocean model lies the conservation of energy. The temporal evolution of the surface layer temperature ($\partial T_\text{s} / \partial t$) at a particular oceanic grid point is governed by the net heat fluxes, expressed as:
\begin{equation}
\label{eqn:general-equation}
\rho C_\text{p} H \frac{\partial T_\text{s}}{\partial t} = F_\text{surf} + F_\text{o} + Q_\text{flux},
\end{equation}
where $\rho$ is the density of seawater, $C_\text{p}$ its specific heat capacity (refer to Table~\ref{tab:ocean-physical-parameters} for associated values) and $H$ the depth of the mixed layer. The left-hand side of Eq.~(\ref{eqn:general-equation}) represents the rate of change of heat content per unit area of the slab ocean. 
The right-hand side includes the net surface flux $F_\text{surf}$ computed by the atmosphere model, an ocean heat flux term $F_\text{o}$ (horizontal and vertical transport) computed by our dynamical slab ocean model and an (optional) additional forcing ($Q_\text{flux}$) that accounts for other processes. The net surface flux is given by $F_\text{surf} = F_\text{rad} + F_\text{sens} + F_\text{lat}$, where $F_\text{rad}$ includes net shortwave and longwave radiation, $F_\text{sens}$ represents enthalpy exchange due to temperature differences between the ocean surface and the air, and $F_\text{lat}$ captures latent heat exchange during phase changes of water. 
The $Q_\text{flux}$ term can be used in the absence of $F_\text{o}$ to approximate the effects of ocean mixing and dynamics, but it can also represent other processes depending on the simulated planet (e.g., geothermal heating, important in exoplanet contexts). A positive (negative) sum of $F_\text{surf} + F_o + Q_\text{flux}$ leads to a net gain (loss) of heat, increasing (decreasing) the ocean surface temperature.


\subsection{\label{subsec:sea-ice-evolution}Sea ice evolution}

Sea ice develops when the ocean temperature falls below the freezing point of seawater ($T_\text{freeze} =$ --1.8$^\circ\mathrm{C}$)\footnote{This value reflects modern Earth's average ocean salinity of 35 psu. The model's $T_\text{freeze}$ value can be changed as per the seawater salinity of the model planet.}. The vertical temperature profile within the sea ice linearly varies between a fixed bottom value $T_\text{freeze}$ in contact with the ocean, and a prognostic surface ice temperature $T_i$. The mean temperature of the ice is therefore equal to $(T_i+T_\text{freeze})/2$. A separate snow layer with its own temperature, thickness and finite heat capacity can accumulate on top of the ice. 

The evolution of ice extent and thickness is governed by energy conservation during phase changes, ensuring that the ocean temperature remains at $T_\text{freeze}$ as long as ice is present. Specifically, if the temperature of the surface ocean layer ($T_\text{s}$) instantaneously falls below its freezing point, its temperature is set back to $T_\text{freeze}$ and the resulting energy difference creates ice.  
Conversely, if the ocean temperature exceeds $T_\text{freeze}$ when ice is present, the temperature is set back to $T_\text{freeze}$, and the excess energy is used to melt part of the ice. 
A similar mechanism governs melting from above. If snow has accumulated on top of the sea ice and the surface temperature (of snow-covered ice) exceeds $T_\text{melt} = 0\,^\circ\mathrm{C}$, it is reset to $T_\text{melt}$ and the energy difference is first used to melt the snow and then the ice. If the heat fluxes are strong enough such that energy is still remaining, it is used to warm the ocean, thus satisfying energy conservation. Sea ice may grow or melt primarily by adjusting either its area or thickness, depending on whether the dominant energy flux is from the ocean below or the atmosphere above.

As described above, snow and ice are both subject to surface heat fluxes from the atmosphere ($F_{\text{a}-\text{i}}$) and conductive fluxes from the ocean and within the ice itself ($F_{\text{i}-\text{o}}$). Ocean heat transport or q-flux changes the ocean temperature, and so, only acts indirectly on the ice. Following \citet{codron2012ekman}, we rewrite Eq.~(\ref{eqn:general-equation}) to specifically describe the evolution of the ice temperature $T_\text{i}$:
\begin{equation}
\label{eqn:ice-equation}
\rho_\text{i} C_\text{i} H\frac{\partial T_\text{i}}{\partial t} = 2 [F_{\text{a}-\text{i}} - F_{\text{i}-\text{o}}]
\end{equation}
where $\rho_\text{i}$ is the ice density, $C_\text{i}$ is its specific heat capacity, and $F_{\text{i}-\text{o}}=\frac{\lambda}{H}(T_\text{i} - T_0)$ is a vertical conductive flux, where $\lambda$ is the thermal conductivity of ice (refer to Table~\ref{tab:ocean-physical-parameters} for associated values). The factor of 2 accounts for the linear (conductive) thermal profile within the ice. If a snow layer is present, we also account for the associated snow-ice conductive flux.


\begin{figure}[h]
\includegraphics[width=15cm]{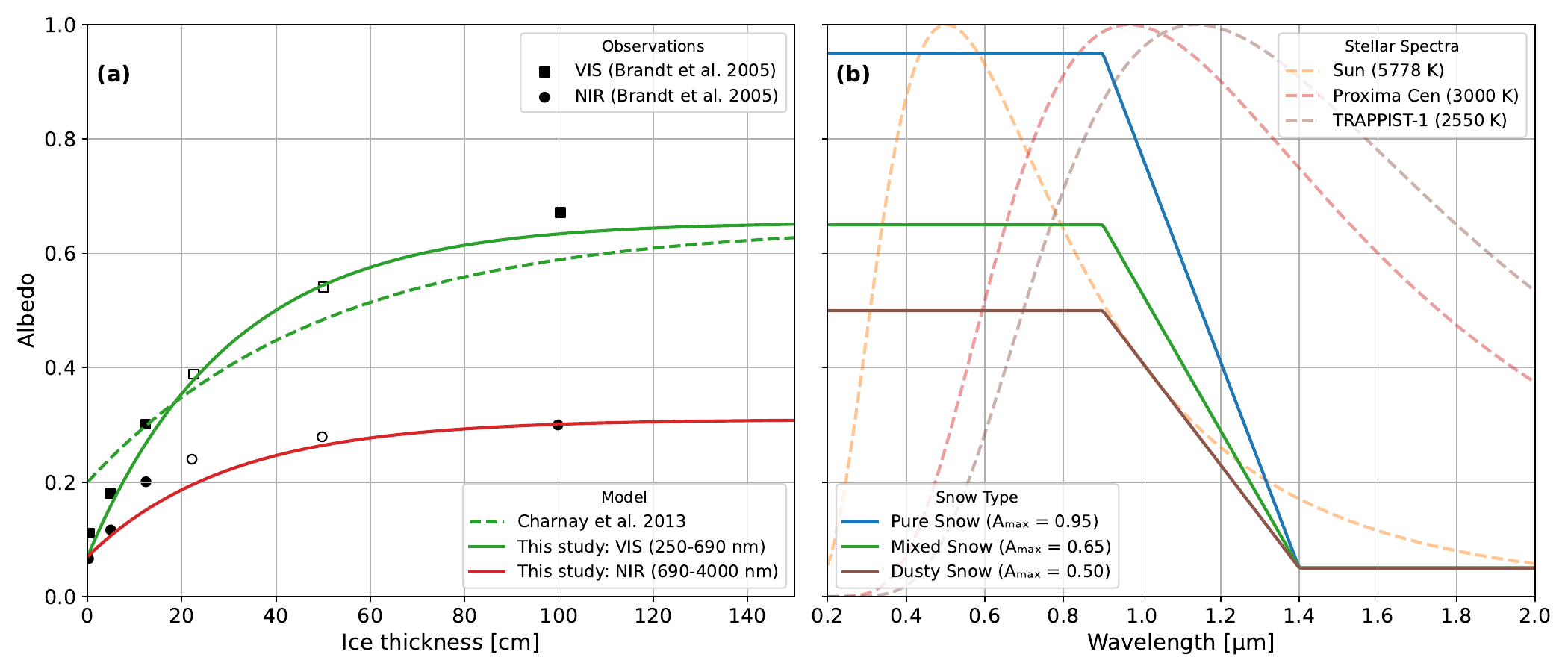}
\vspace{-1em} 
\caption{\label{fig:albedo_parameterisation}\textbf{(a)} Bare sea ice albedo as a function of ice thickness in the Visible (VIS; green) and Near-Infrared (NIR; red) bands in our study, with curves fit to Antarctic observations from \citet[][filled symbols: measurements; open: interpolated]{brandt2005surface}. The spectrally-independent albedo used in \citet{charnay2013exploring} is shown for reference (dashed green).
\textbf{(b)} Broadband spectral distribution of snow and ice albedo for pure (blue), mixed (green), and dusty (brown) snow—with A$_\text{max}$ values of 0.95, 0.65, and 0.50, respectively. Dashed lines indicate the normalised black body emission of the Sun (orange), Proxima Centauri (red), and TRAPPIST-1 (brown).}
\end{figure}

The parameterisation of sea ice and snow albedo also plays a critical role in shaping a model planet's climate and consequently, its observability. \citet{charnay2013exploring} addressed this by treating the albedo of bare sea ice as a function of ice thickness. In their framework, the surface albedo ($A$) at an oceanic grid point could correspond to that of open ocean, overlying snow (if present), or bare sea ice, the latter varying with thickness according to:
\begin{equation}
\label{eqn:albedo-equation}
A = A_{\text{ice}}^\text{max} - (A_{\text{ice}}^\text{max} - A_{\text{ice}}^\text{min}) e^{-h_{\text{ice}}/h_{\text{ice}}^0}
\end{equation}
In the present study, we improve this scheme by incorporating a spectral dependence of albedo in addition to the existing thickness dependence. Figure~\ref{fig:albedo_parameterisation}a shows how the model's prescribed sea ice albedo varies as a function of ice thickness in the Visible (VIS; 250-690 nm; green line) and Near-Infrared (NIR; 690-4000 nm; red line) spectral bands. The spectrally-independent albedo profile from \citet{charnay2013exploring} is also shown for reference (dashed green).
This earlier scheme assigned an albedo of 0.2 to newly formed ice, whereas our formulation introduces a physically motivated transition, with the albedo increasing smoothly from its minimum possible value, i.e., $A_{\text{ice}}^\text{min} \equiv A_\text{ocean} = 0.07$,\footnote{The albedo of open water primarily results from Fresnel reflection and is largely wavelength-independent, consistent with measurements \citep[e.g.,][]{payne1972albedo,katsaros1985albedo,pegau2001albedo}.} up to a maximum value $A_{\text{ice}}^\text{max}$, following Eq.~(\ref{eqn:albedo-equation}).
We calibrate the maximum bare sea ice albedo $A_{\text{ice}}^\text{max}$ by fitting this expression to long-term observations of Antarctic sea ice albedo, as documented by \citet[][filled symbols in Fig.\ref{fig:albedo_parameterisation}]{brandt2005surface}, and summarised by \citet{pedersen2009new}.
From Table 3 in \citet{brandt2005surface}, the VIS (NIR) albedos of bare first year sea ice typically range from $\approx0.54$ (0.27) for thinner ice up to $\approx0.67$ (0.31) for thicker, cold, snow-free ice. Fitting Eq.~\ref{eqn:albedo-equation} to these data yields $A_{\text{ice}}^\text{max} \approx 0.65$ in the VIS and $\approx 0.31$ in the NIR. The parameter $h_{\text{ice}}$ denotes sea ice thickness (in m), while $h_{\text{ice}}^0$ is a scaling parameter (set to 0.3 m) that governs the rate of transition between the minimum and maximum albedo values. Since our formulation differs from that of \citet{charnay2013exploring} in both the minimum albedo (now fixed to the ocean albedo) and in its spectral dependence, the original value of $h_{\text{ice}}^0 = 0.5$ m is no longer optimal. We therefore refit Eq.~(\ref{eqn:albedo-equation}) to the \citet{brandt2005surface} observations, yielding $h_{\text{ice}}^0 = 0.3$ m. The older would still preserve the maximum albedo for thick ice, but would systematically underestimate albedos at intermediate thicknesses.

Figure~\ref{fig:albedo_parameterisation}b shows the Generic-PCM's broadband spectral distribution of snow and ice albedo for three idealised surface cases: pure (blue line), mixed (green), and dusty (brown) snow. These are based on the works of \citet{warren1980model}, \citet{warren1984impurities} and \citet{joshi2012suppression} and are characterized by maximum albedos ($A_\mathrm{max}$) of 0.95, 0.65, and 0.50, respectively. Currently, our model does not account for an age-dependent snow albedo or impurity accumulation. 
We therefore adopt the mixed snow spectral albedo profile, for both snow on land and for snow-covered sea ice, following \citet{turbet2016habitability}. The corresponding value for Sun-like stars (0.65) is consistent with the mean of the snow and ice albedos (0.8 and 0.5 respectively) reported in \citet{joshi2012suppression}. The mixed snow profile also represents a physically reasonable intermediate state between fresh clean snow and old dusty snow, and can be interpreted as a climatological mean snow albedo. The pure snow profile would overestimate the strength of the ice-albedo feedback, while the dusty snow profile would underestimate it (see Fig.~\ref{fig:earth-zon-avg} for more details). 
The dashed lines in Fig.~\ref{fig:albedo_parameterisation}b indicate the normalised black body emission of the Sun (orange, 5778 K), Proxima Centauri (red, 3000 K), and TRAPPIST-1 (brown, 2550 K). For these stars, the bolometric albedo of mixed snow is approximately 0.55, 0.36 and 0.29 respectively\footnote{Assuming blackbody emission is a weak assumption for M-dwarfs, which exhibit significant absorption features.}, highlighting the importance of including a spectral energy distribution in determining surface energy budgets.

\subsection{\label{subsec:heat-transport}Heat transport by ocean circulation}
\begin{figure}[ht]
\includegraphics[width=18cm]{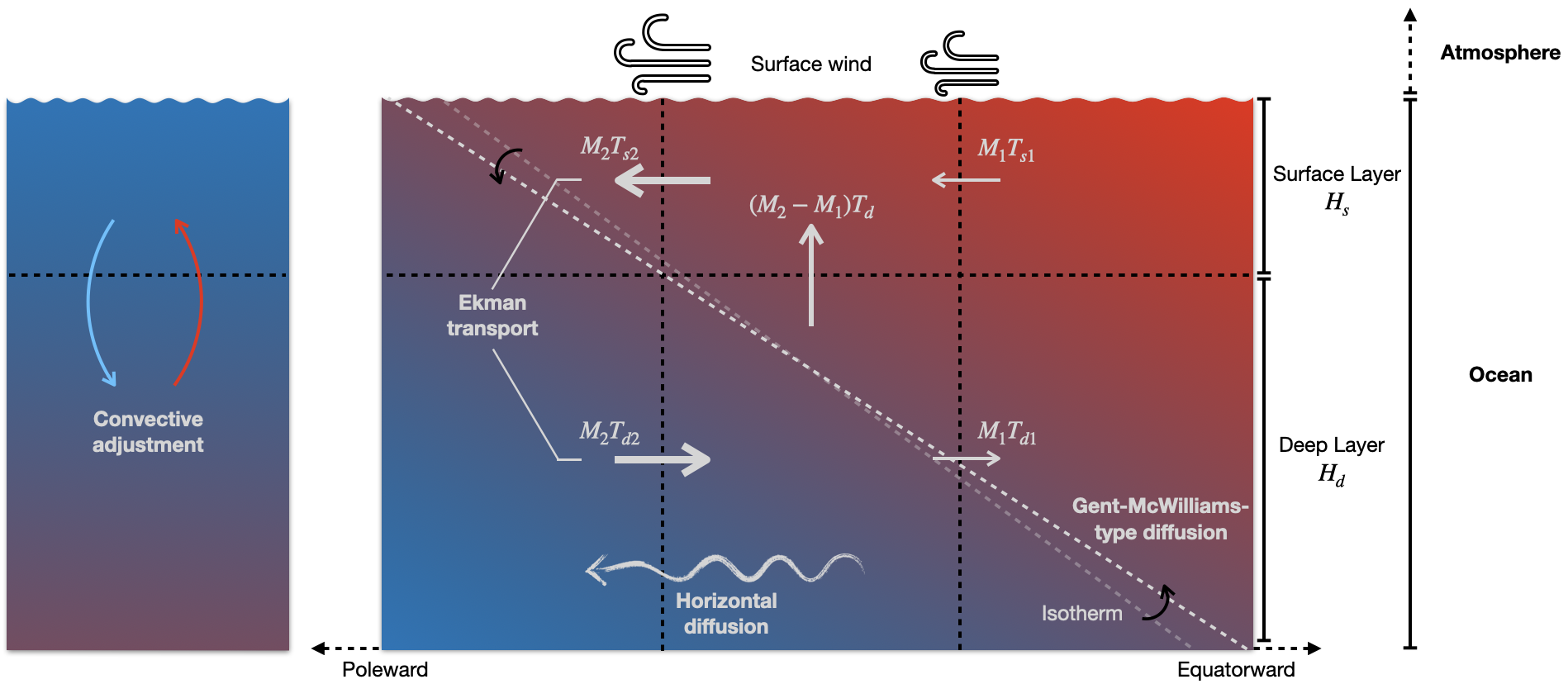}
\vspace{-1em} 
\caption{\label{fig:OHT_Diagram}Cross-sectional representation of the ocean along latitude and depth. The top section represents the oceanic mixed (surface) layer and the bottom represents the deep ocean layer. The different heat transport regimes are visible: the wind-driven Ekman transport and eddy advection based on the Gent-McWilliams (GM) scheme both yield horizontal mass fluxes across the grid interface ($M_1, M_2$) opposed in the top and bottom layers. These are proportional to the surface wind stress (Ekman) or isotherm slope (GM). The vertical mass flux is deduced from mass conservation. The horizontal diffusive flux (wavy arrow) is downgradient in both layers. Convective adjustment, particularly relevant for high-latitudes, is represented on the left.}
\end{figure}

The nomenclature "dynamical slab ocean model" is purposefully chosen to reflect our model's emphasis on ocean dynamics.
Figure~\ref{fig:OHT_Diagram} provides a cross-sectional view of the ocean model in the horizontal (latitude) and vertical (depth) dimensions, and illustrates the model's various OHT regimes. The upper section depicts the oceanic mixed (surface) layer, with a fixed thickness of $H_s =$ 50 m, where interactions with the atmosphere occur (see Sect.~\ref{subsec:model-limitations} for a discussion of the implications of using a fixed mixed-layer depth). The lower section corresponds to the deep ocean, with a thickness of $H_d =$ 150 m, which exchanges heat only with the surface layer.

In our model, heat transport by the ocean is represented by four components, detailed through Sects.~\ref{subsubsec:ekman-transport}--\ref{subsubsec:convective-adjustment}:

\subsubsection{\label{subsubsec:ekman-transport}Wind-driven Ekman transport}

We closely follow the 2-Layer Ekman model detailed in \citet{codron2012ekman}. A wind-driven horizontal mass flux $M$ integrated over the ocean surface mixed layer is first computed as
\begin{equation}
  \left\{
  \begin{aligned}
    M_x&=&(\epsilon\tau_x+f\tau_y)/(\epsilon^2+f^2)\\
    M_y&=&(\epsilon\tau_y-f\tau_x)/(\epsilon^2+f^2)
  \end{aligned}
  \right.
  \label{eqn:massflux}
\end{equation}
where $M_x$ and $M_y$ are the zonal and meridional oceanic mass fluxes and $\tau_x$ and $\tau_y$ are the corresponding wind stress components, $f$ is the Coriolis parameter ($f=2\Omega \sin{\phi}$, where $\Omega$ is the rotation rate of the planet and $\phi$ the latitude) and $\epsilon$ is a frictional damping coefficient. In the limit when \( |f| \gg \epsilon \), the wind-driven flux becomes equal to the Ekman transport $\mathbf{M_E} = - \frac{1}{f} \mathbf{k} \times \mathbf{\tau}$, where $\mathbf{k}$ is the unit vector in the vertical direction. These mass fluxes are then used to advect the surface layer temperature, while equal and opposite mass fluxes are applied in the deep layer to conserve mass.

This process is illustrated on Fig.~\ref{fig:OHT_Diagram} along a single horizontal dimension, with frictional mass fluxes $M_1$ and $M_2$ at two neighbouring grid point boundaries in the surface layer and opposite fluxes at depth. If these horizontal mass fluxes are divergent -- as on Fig.~\ref{fig:OHT_Diagram} -- they give rise to compensating upwelling mass fluxes (or downwelling, when there is surface convergence), that advect the temperature vertically. The OHT term in the surface layer temperature evolution is then given by the divergence of heat fluxes:
\begin{equation}
\label{eqn:ekman-equation}
\rho H_s S \frac{\partial T_s}{\partial t} = M_1T_{s1} - M_2T_{s2}+(M_2-M_1)T_d
\end{equation}
Here, $M_1$ is the vertically integrated mass flux computed from the wind stress for the latitude closer to the equator and $M_2$ is that for the latitude closer to the pole (for the meridional component of the transport). Similarly, $T_{s1,2}$ are the temperatures of the surface layer at the grid interfaces, computed using an upstream scheme. $T_d$ is the temperature of the deep (upwelled) layer in the example shown, but is replaced by the surface temperature $T_s$ in the case of downwelling. $S$ is the horizontal area of the grid point. The temperature of the deep layer evolves in the same manner, but with opposite signs for the mass transports.

Our study also includes a modification of this scheme near the equator where \( f \rightarrow 0 \). According to Eq.~(\ref{eqn:massflux}), the mass transport becomes purely frictional and aligned with the wind in this limit. While this is indeed observed in the surface layers of Earth's oceans, the meridional mass transport component responsible for most of the energy transport across the equator on Earth is better captured by the Sverdrup balance, given by \( M_\text{Sv} = \frac{1}{\beta} \mathbf{k} \cdot (\nabla \times \boldsymbol{\tau}) \), where \( \beta = \left. \frac{\partial f}{\partial \phi} \right|_{\phi \rightarrow 0^\circ} \). We therefore include a smooth transition from the frictional Ekman balance to a curl-driven Sverdrup balance near the equator. This transition is handled using a weighting function of the form \( \exp(-f^2/\epsilon^2) \), so that Sverdrup transport dominates when \( |f| \lesssim \epsilon \), and is negligible at higher latitudes where \( |f| \gg \epsilon \). For Earth, with \( \epsilon = 10^{-5} \,\mathrm{s}^{-1} \), this corresponds to a transition centered around \( \pm 4^\circ \) latitude. This addition improves the modelled meridional transport in the tropics (see Sects.~\ref{subsec:gm-discussions} and \ref{subsec:hemisphere-asymmetry}), especially on Earth-like rotating planets. However, this approach may have to be modified for slowly rotating planets:  \( f \) and \( \beta \) become small, and the Sverdrup transport would become unrealistically large compared to the direct frictional transport.

\subsubsection{\label{subsubsec:gent-mcwilliams}Gent-McWilliams transport}

We model mesoscale ocean eddies using the Gent–McWilliams (GM) parameterisation \citep{gent1990isopycnal}, which represents the effect of eddies along isopycnals (surfaces of constant density). In the full theory, a residual eddy-induced transport velocity -- with both horizontal and vertical components -- is derived from the large-scale density field, with an overturning streamfunction proportional to the slope of the isopycnals \citep[see Eq. 2 in][]{danabasoglu1995sensitivity}. However, since our model does not include salinity, density reduces to a function of temperature alone, and so, the isopycnals are equivalent to the isotherms. 

Like in the Ekman transport case, horizontal mass fluxes are computed at each grid interface, with opposite values in the surface and deep layers. This is equivalent to evaluating the GM streamfunction only at the interface between the two layers.
The horizontal mass flux $\textbf{M}_\text{GM}$ in the surface layer is then given by $\textbf{M}_\text{GM}=\rho\kappa_\text{GM}\boldsymbol{\sigma}$, where $\kappa_\text{GM}$ is the GM thickness diffusivity coefficient and $\boldsymbol{\sigma}$ denotes the isotherm slope. These horizontal fluxes, along with vertical fluxes deduced from mass conservation, are then used to advect temperature. The amplitude of the eddy-induced mass transport is therefore proportional to the local isotherm slope, and acts to flatten tilted isotherms by transporting fluid along them -- from deeper to shallower regions. This is illustrated in Fig.~\ref{fig:OHT_Diagram}, where a representative isotherm is shown as a white dashed line.

We prescribe a GM diffusion coefficient of 2000 m$^2$s$^{-1}$ and a maximum slope of 0.002 for numerical stability. These values were tuned to fit the meridional ocean heat transport profile of an Earth-like aquaplanet using the MITgcm \citep{marshall1997finite,marshall1997hydrostatic,adcroft2004implementation}, an AOGCM (see Sects.~\ref{subsec:zonal-average-aquaplanet} and \ref{subsec:gm-discussions}).
The eddy-induced transport acts to reduce the slope of the isotherms, shown in Fig.~\ref{fig:OHT_Diagram} as the grey dashed line transforming into the white dashed line, resulting in high-latitude surface heating and low-latitude depth cooling. The vertically integrated eddy transport acts to redistribute heat from regions of higher to lower temperature, effectively behaving as a downgradient energy flux; but there is also a re-stratification of the ocean, increasing the temperature difference between the surface and deep ocean, in turn making the Ekman transport larger according to Eq.~(\ref{eqn:ekman-equation}). In contrast, a purely diffusive flux would transport heat towards higher latitudes for both the surface and deep layers. Importantly, we also find that this restratification can simulate the effect of local convection (when the surface becomes colder than the deep ocean) quite well (see Sect.~\ref{subsec:gm-discussions}).

\subsubsection{\label{subsubsec:horizontal-diffusion}Horizontal diffusion}

Mixing caused by turbulence and subgrid-scale eddies can also be accounted for by horizontal eddy diffusion (represented with a wavy arrow in Fig.~\ref{fig:OHT_Diagram}), with a uniform diffusivity in both ocean layers. \citet{codron2012ekman} prescribed a horizontal diffusion coefficient value of $D=$ 25000 m$^2$s$^{-1}$ to best reproduce the meridional heat transport of aquaplanets simulated using fully coupled AOGCMs. However, now that we explicitly account for mesoscale eddies through the GM scheme, we prescribe a decreased diffusion coefficient value of 8000 m$^2$s$^{-1}$ so as not to overestimate overall diffusive heating. Similar to the GM coefficient, we tuned this horizontal diffusion coefficient to fit the meridional OHT of an Earth-like aquaplanet, as modelled by the MITgcm \citep{marshall2007mean}. These coefficient values may change as a function of the rotation rate of the planet. We briefly address this in Sect.~\ref{subsec:model-limitations}, while a more detailed investigation is planned for future work.

\subsubsection{\label{subsubsec:convective-adjustment}Convective adjustment}

The two-layer nature of the ocean model also facilitates a convective adjustment. Specifically, if the surface layer becomes colder than the deep layer (for e.g., in winter at mid-high latitudes), vertical convective adjustment ensures that the heat stored at depth is restored to the surface (see the high-latitude slab in the extreme left of Fig.~\ref{fig:OHT_Diagram}). This effectively simulates (denser) colder water at the surface descending and ensures that the deep ocean temperature remains lower than the surface.

\subsubsection{\label{subsubsec:the-equation}The Dynamical Slab Ocean Equation}

We now consolidate these OHT processes into the governing equations for the evolution of the surface and deep layer temperatures (denoted by $T_s$ and $T_d$ respectively) at a given grid point. The evolution of the surface layer temperature is expressed as: 
\begin{equation}
\label{eqn:modified-Ts-evolution}
\frac{\partial T_s}{\partial t} = \frac{1}{\rho C H_s} (F_{a-o} + F_{i-o} + F_c) + \text{D} \nabla^2 T_s - \frac{1}{\rho H_s} \left[ \operatorname{div}_H \left[ (\textbf{M}_{\text{Ek}} + \textbf{M}_{\text{GM}}) T_s \right] - \left( \textbf{W}_{\text{Ek}} + \textbf{W}_{\text{GM}} \right) \hat{T} \right],
\end{equation}
where, $F_{a-o}$ and $F_{i-o}$ represent the heat fluxes from the atmosphere and the sea ice to the ocean respectively, and $F_c$ denotes the \textbf{convective heat flux} between the two slab layers. $\text{D}$  is the \textbf{horizontal diffusion} coefficients. The horizontal divergence operator, $\operatorname{div}_H$, calculates the net horizontal heat inflow or outflow at a grid point. \textbf{Ekman transport} and the \textbf{GM scheme} both change the horizontal and vertical mass fluxes. We decompose the horizontal mass flux, \textbf{M}, into its Ekman and GM contributions as \(\textbf{M} = \textbf{M}_{\text{Ek}} + \textbf{M}_{\text{GM}}\). The horizontal mass flux, \textbf{M} (with units of kg m$^{-1}$ s$^{-1}$), along with the advected surface and deep temperatures $T_s$ and $T_d$, are evaluated on the grid-point interfaces. The vertical mass flux, \textbf{W}, can also be decomposed as \(\textbf{W} = \textbf{W}_{\text{Ek}} + \textbf{W}_{\text{GM}}=\operatorname{div}_H \textbf{M}_{\text{Ek}}+\operatorname{div}_H \textbf{M}_{\text{GM}}\). The vertically-advected temperature is denoted by $\hat{T}$. If $\textbf{W}>0$, water from depth is upwelled and $\hat{T}=T_d$ and if $\textbf{W}<0$, surface water is downwelled and $\hat{T}=T_s$.
 
The deep layer temperature, $T_d$, evolves in a similar way to the surface layer but without the direct influence of surface fluxes, i.e., $F_{a-o}$ and $F_{i-o}$. It changes due to convection, diffusion, vertical and horizontal advection. We use the same Ekman and GM transports as in the surface layer, but with a negative sign to reflect the opposing direction of the return flow. The evolution of $T_d$ at a given grid point is then governed by:
\begin{equation}
\label{eqn:modified-Td-evolution}
\frac{\partial T_d}{\partial t} = \frac{1}{\rho C H_d} F_c  
+\text{D} \nabla^2 T_d + \frac{1}{\rho H_d} \left[ \operatorname{div}_H \left[ (\textbf{M}_{\text{Ek}} + \textbf{M}_{\text{GM}}) T_d \right] 
- \left( \textbf{W}_{\text{Ek}} + \textbf{W}_{\text{GM}} \right) \hat{T} \right]
\end{equation}




Our model does not include other pressure-driven ocean circulation features such as salinity-driven circulations or geostrophic currents such as horizontal gyres. Despite this, we find that our model reproduces the global meridional OHT quite closely compared to a full AOGCM, as we show in Sects.~\ref{sec:aquaplanet} and~\ref{sec:earth}.

\section{\label{sec:aquaplanet}Model validation with a coupled aquaplanet}

\begin{table}[h!]
\centering
\begin{tabular}{|c|c|c|}
\hline
\textbf{Physical Parameter} & \textbf{Aquaplanet} & \textbf{Earth} \\
\hline
Stellar spectrum & \multicolumn{2}{c|}{Sun} \\
\hline
Solar constant (W/m$^2$) & \multicolumn{2}{c|}{1366} \\
\hline
Rotation period (h) & \multicolumn{2}{c|}{24} \\
\hline
Orbital period (d) & \multicolumn{2}{c|}{365} \\
\hline
Radius (km) & \multicolumn{2}{c|}{6378.137} \\
\hline
Surface gravity (m/s$^2$) & \multicolumn{2}{c|}{9.81} \\
\hline
Surface roughness coefficient (m) & \multicolumn{2}{c|}{0.01} \\
\hline
H$_2$O cloud droplet radius ($\mu$m) & \multicolumn{2}{c|}{12 (liquid) and 35 (ice)$^\dagger$} \\
\hline
Average surface pressure (bar) & \multicolumn{2}{c|}{1.013} \\
\hline
N$_2$ partial pressure (bar) & \multicolumn{2}{c|}{0.9996} \\
\hline
H$_2$O partial pressure (bar) & \multicolumn{2}{c|}{Variable} \\
\hline
CO$_2$ partial pressure (mbar) & \multicolumn{2}{c|}{0.375 (370 ppm)$^\dagger$} \\
\hline
Obliquity ($^\circ$) & 0.0 or 23.44 & 23.44\\
\hline
Eccentricity & 0.0 & 0.0167\\
\hline
Topography & flat & Earth's continents$^\dagger$\\
\hline
Surface albedo & 0.07 & Earth's albedo$^\dagger$\\
\hline
Thermal inertia (J/m$^2$/s$^{1/2}$/K) & 18000 & Earth's thermal inertia$^\dagger$\\
\hline
\end{tabular}
\begin{flushleft}
$^\dagger$ Values adopted from \cite{charnay2013exploring}.
\end{flushleft}
\caption{Summary of the main planetary parameters used in the GCM for the study.}
\label{tab:physical-parameters}
\end{table}

To clarify the OHT mechanisms, we first model an Earth-like atmosphere-ocean coupled aquaplanet — an idealised planet with 100\% surface ocean coverage. The main physical parameters we use for the simulation are presented in Table~\ref{tab:physical-parameters}. In our first set of aquaplanet simulations, we set both obliquity and eccentricity to 0 to eliminate orbital-parameter-induced temporal asymmetries. The atmosphere is composed of N$_2$ with CO$_2$ at close to modern Earth levels (0.375 mbar) and the initial average surface pressure is 1013 mbar.

We conduct a suite of simulations, under 7 different OHT regimes, detailed in Table~\ref{tab:aquaplanet-experiments} to assess their individual impacts on the planet's climate. Each simulation is initialised with an isothermal atmosphere and a uniform sea surface temperature (SST) of 290 K. The model typically achieves a steady state within 20 years\footnote{At the spatiotemporal resolution used in this study, one model year requires approximately 2.75 hours of computation on 24 cores.} when launched from an idealised meridional temperature profile (as in \citealt{codron2012ekman}; similar to the QOBS profile from \citealt{neale2000standard}), and within 40 years when initialised from an isothermal 290 K state. We also validate the performance of our dynamical slab ocean model compared to a full AOGCM, the MITgcm, specifically comparing our results with those described in \citet{marshall2007mean} and \citet{brunetti2019co}.

\begin{table}[h!]
\centering
\begin{tabular}{|c|c|c|c|c|}
\hline
\textbf{Case} & \textbf{Horizontal diffusion} & \textbf{Gent-McWilliams} & \textbf{Ekman} & \textbf{Convective adjustment} \\
\hline
1 & -- & -- & -- & -- \\
\hline
2 & + & -- & -- & -- \\
\hline
3 & -- & -- & + & -- \\
\hline
4 & + & -- & + & -- \\
\hline
5 & -- & + & + & -- \\
\hline
6 & + & + & + & -- \\
\hline
7 & + & + & + & + \\
\hline
\end{tabular}
\vspace{1em} 
\caption{Table showing the aquaplanet simulations tested using the dynamical slab ocean model. Case 1 represents a slab ocean with all oceanic heat transport (OHT) mechanisms turned off. Each successive case activates specific OHT processes, culminating in Case 7, which includes all OHT mechanisms: horizontal diffusion, Gent-McWilliams advection, Ekman transport and convective adjustment.}
\label{tab:aquaplanet-experiments}
\end{table}

\subsection{\label{subsec:zonal-average-aquaplanet}Meridional climatic features}

\begin{figure}[h]
\centering
\includegraphics[width=12cm]{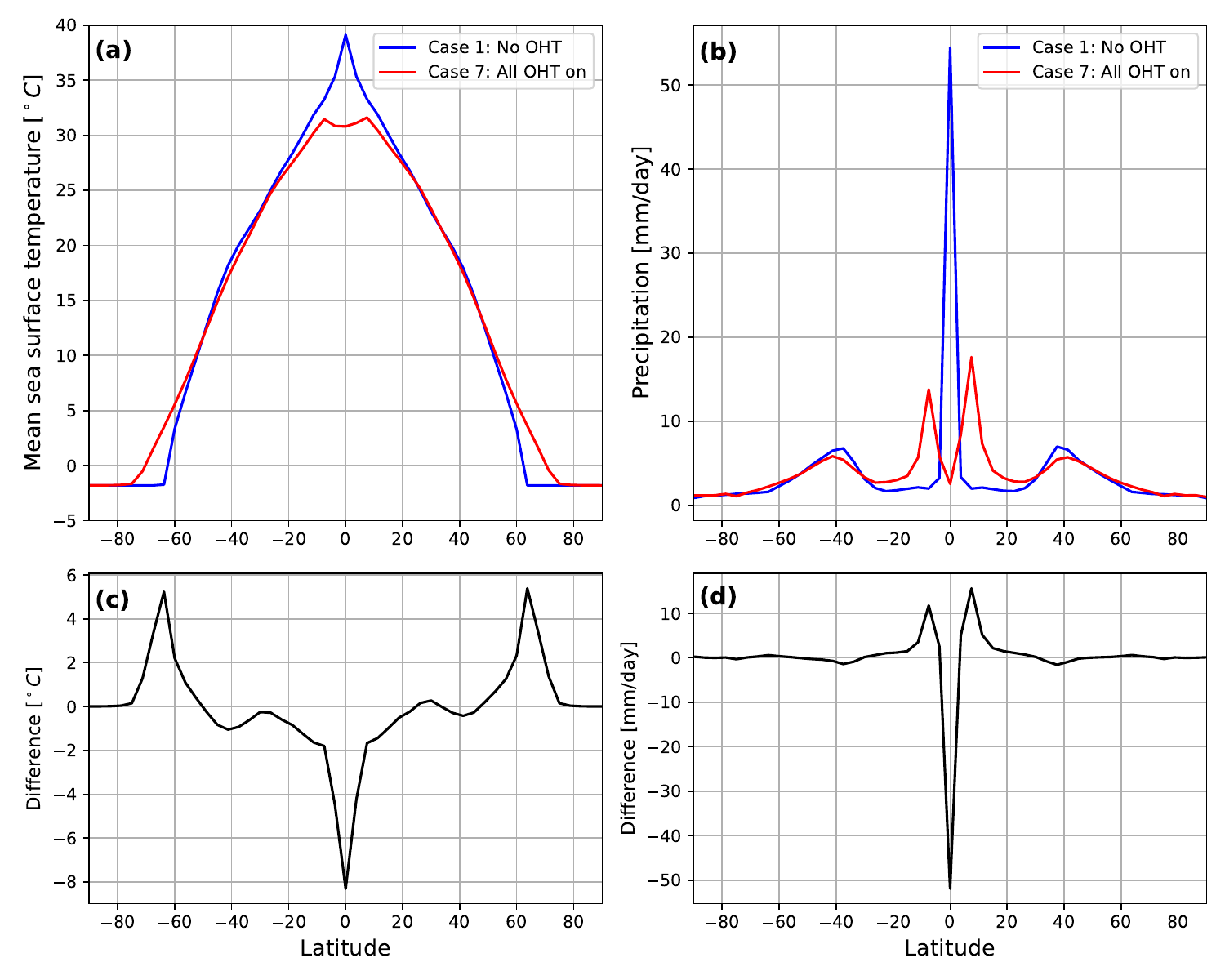}
\vspace{-1em} 
\caption{\label{fig:zonal-temperature-aquaplanet}\textbf{(a)} Zonal mean sea surface temperature (SST, 10-year average) of the aquaplanet for Case 1 (no oceanic heat transport, OHT) and Case 7 (all OHT on) simulations. \textbf{(c)} Difference in SST (ON -- OFF), showing the effect of OHT on the same. \textbf{(b)} Zonal mean precipitation (rain + snow, 10-year average) for the same simulations. \textbf{(d)} Difference in precipitation (ON -- OFF)}
\end{figure}

Figure~\ref{fig:zonal-temperature-aquaplanet}a shows the decadal-averaged zonal mean SST profile, for the two extreme configurations: Case 1 (no OHT) and Case 7 (all OHT mechanisms enabled) experiments. The corresponding SST difference is shown in Fig.~\ref{fig:zonal-temperature-aquaplanet}c, highlighting the impact of OHT. In Case 1 (Fig.~\ref{fig:zonal-temperature-aquaplanet}a, blue line), the SST profile peaks sharply at the equator as there is no mechanism to redistribute heat from this location, with additional warming due to enhanced moisture and high cloud coverage.
The inclusion of all OHT processes (Case 7, red profile) brings two important distinctions. Firstly, the equatorial SST is cooler by $\approx$ 8$^\circ$C (see Fig.~\ref{fig:zonal-temperature-aquaplanet}c) compared to Case 1, with a flatter temperature profile and the presence of an equatorial cold tongue. This is driven by Ekman transport, which causes cold water upwelling due to the strong divergence of heat transport at the equator. Additionally, the polar regions are warmer (see Fig.~\ref{fig:zonal-temperature-aquaplanet}c) due to diffusion (both, large scale eddy diffusivity and the GM scheme), pushing the ice line poleward by 10-15$^\circ$ in latitude. The retreating ice line latitude further strengthens the warming of the planet through the ice-albedo-feedback. These features are also evident in the annually averaged SST maps of Case 1 and Case 7 shown in Fig.~\ref{fig:mixed-layer-temperature-aquaplanet}, as well as in the corresponding animations (see the Video supplement section). Figure~\ref{fig:zonal-temperature-aquaplanet}c also clearly shows the regions of (Ekman-induced) equatorial cooling and (diffusion and GM-driven) polar warming.

The corresponding zonal mean precipitation profiles (rain + snow) for Cases 1 and 7 are shown in Fig.~\ref{fig:zonal-temperature-aquaplanet}b, and the precipitation difference (OHT on -- OHT off) in Fig.~\ref{fig:zonal-temperature-aquaplanet}d. The high equatorial temperature observed in Fig.~\ref{fig:zonal-temperature-aquaplanet}a for Case 1 (OHT off) corresponds to strong water vapour convergence, resulting in the pronounced peak in equatorial precipitation (blue profile) seen in Fig.~\ref{fig:zonal-temperature-aquaplanet}b. On the other hand, in Case 7, equatorial upwelling of cold water prevents precipitation at the equator. This leads to a double-banded precipitation pattern in the tropics on either side of the equator (Fig.~\ref{fig:zonal-temperature-aquaplanet}b), effectively creating a double Intertropical Convergence Zone (ITCZ). The slight hemispheric asymmetry in Fig.~\ref{fig:zonal-temperature-aquaplanet}b may be due to amplifications in internal feedbacks and numerical artefacts. This asymmetry is also visible in the related atmospheric profiles (see Fig.~\ref{fig:aquaplanet-atmosphere-profiles}). Direct overlays of the SST and precipitation profiles of \citet{codron2012ekman} and a related discussion are presented in Appendix~\ref{subsec:direct-overlays-codron-sst-prec}.

\begin{figure}[h]
\centering
\includegraphics[width=10cm]{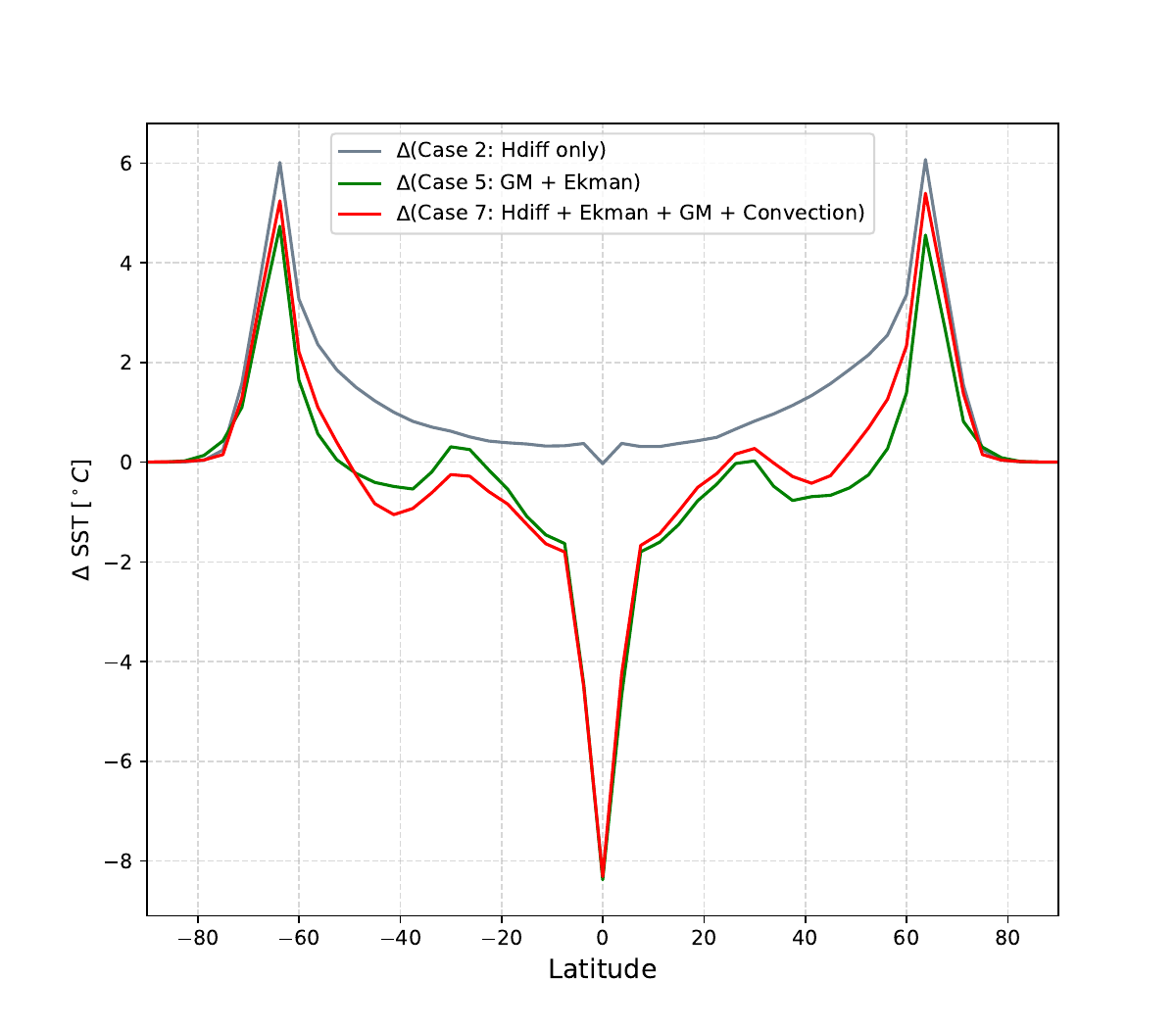}
\vspace{-1em} 
\caption{\label{fig:all-scenarios-sst-difference}Zonal-mean sea surface temperature (SST) differences of the aquaplanet between selected simulations and Case 1 (no ocean heat transport, OHT), highlighting the impact of various transport mechanisms on the climate. The grey line shows the SST difference for Case 2 (horizontal diffusion only), the green for Case 5 (Gent–McWilliams and Ekman transport), and the red for Case 7 (all OHT enabled).}
\end{figure}

Figure~\ref{fig:all-scenarios-sst-difference} presents the decadal-mean difference in zonal-mean SST between selected cases from Table~\ref{tab:aquaplanet-experiments} and the baseline Case 1 (no OHT), illustrating how individual OHT mechanisms affect the planet’s climate. In Case 2 (horizontal diffusion only; grey line), SSTs progressively warm with latitude relative to Case 1, reaching +5–6$^\circ$C at mid-to-high latitudes. This highlights the latitudinal smoothing effect of diffusion, with the magnitude of warming dependent on the chosen horizontal diffusion coefficient $D$, and on the mean temperature structure (see the Laplacian term in Eq.~\ref{eqn:modified-Ts-evolution}). In Case 5 (GM + Ekman; green line), the SST profile exhibits strong equatorial cooling (8$^\circ$C) from Ekman-induced upwelling, coupled with moderate warming at subtropical (+2$^\circ$C) and polar (+5$^\circ$C) latitudes. Cases 6 (horizontal diffusion + Ekman + GM) and 7 (all OHT mechanisms active, i.e., Case 6 + convection) are nearly indistinguishable, suggesting that convective adjustment has only a minor effect (<1$^\circ$C across all latitudes), well within model uncertainty. This minimal influence likely results from the GM scheme’s ability to capture vertical restratification effectively (see Sect.~\ref{subsec:gm-discussions}). 
Comparing Case 5 (green line) with Case 7 (red line) reveals that the addition of horizontal diffusion primarily enhances mid-latitude warming. Lastly, simulations with Ekman transport display minor hemispheric asymmetries. Since Ekman transport is wind-driven, these asymmetries may stem from the atmospheric model and warrant further investigation.

The contribution of each oceanic heating term can also be assessed by examining the meridional ocean heat transport. This diagnostic represents how energy is redistributed from regions of surplus (typically the  equator) to those of deficit (poles), offering key insights into the mechanisms driving heat transport in a particular fluid (for e.g., the ocean and/or the atmosphere) as detailed in \cite{marshall2007mean}.  Our model outputs the heating rates (in W/m$^2$) of each oceanic process contributing to the total oceanic heating (i.e., horizontal diffusion, Ekman and GM transport). The heating rate for each latitude can be converted to Northward oceanic heat transport by integrating the heating rate over the surface area of the planet and considering the cumulative transport northward. The northward heat transport $Q$ at latitude $\phi$ is then:
\begin{equation}
\label{eqn:northward-transport}
Q(\phi) = 2 \pi R^2 \cos(\phi) \int_{-\pi/2}^{\phi} \overline{q}(\phi') d\phi',
\end{equation}
where $R$ is the planetary radius and $\overline{q}(\phi')$ the zonal mean heating rate at latitude $\phi'$. 

\begin{figure}[h]
\centering
\includegraphics[width=11cm]{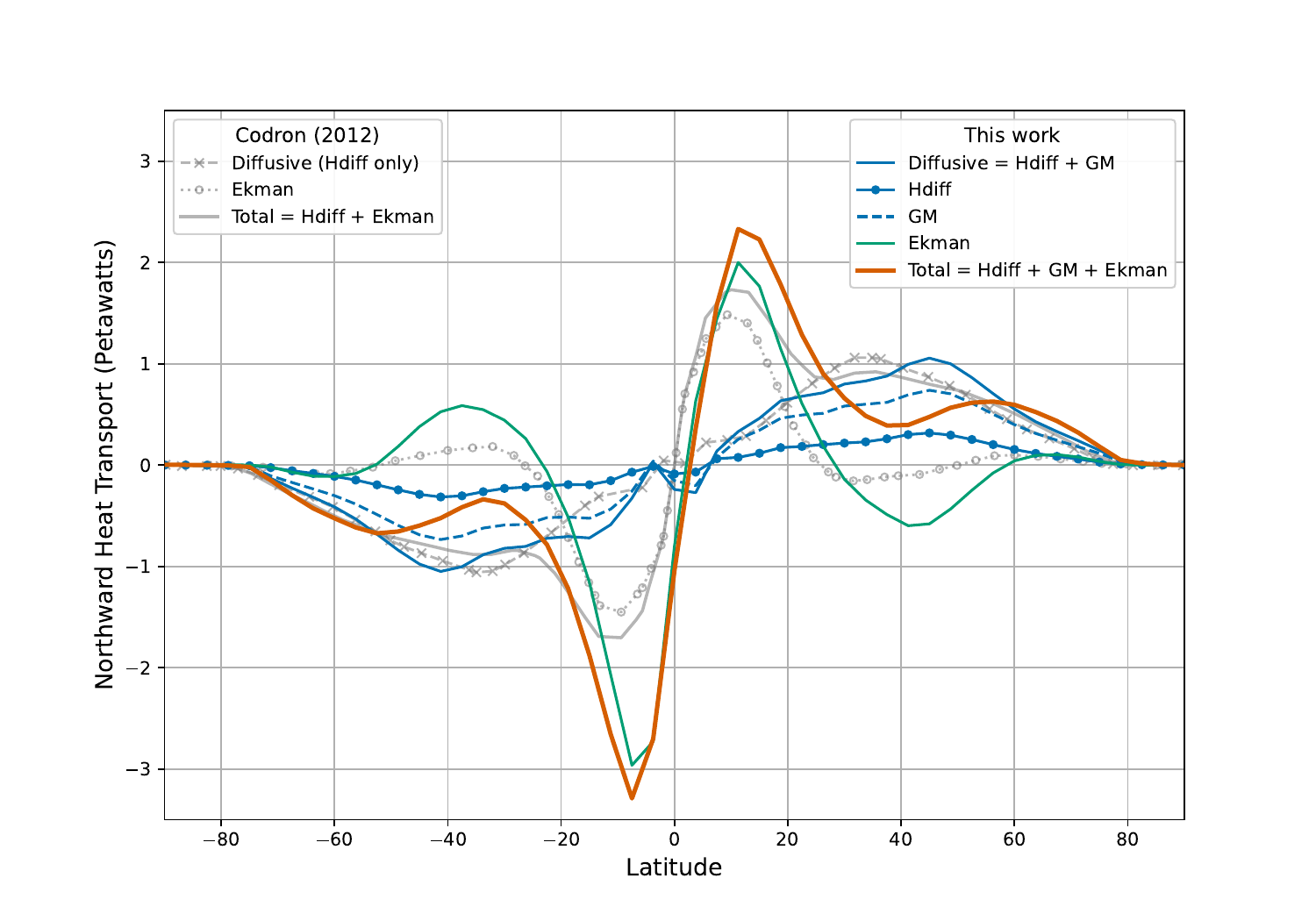}
\vspace{-1em} 
\caption{\label{fig:northward-transport-aquaplanet}Contribution of each ocean heat transport (OHT) term towards the northward meridional OHT (in Petawatts, red line) for Case 7 (all OHT on) of the oblique aquaplanet simulation, with comparison to \citet{codron2012ekman}. 
The decomposed OHT components from our work are shown as coloured lines (blue and green), while those from \citet{codron2012ekman} are plotted in grey.}
\end{figure}

We perform this analysis in the case of an oblique (23.44$^\circ$) aquaplanet. Figure~\ref{fig:northward-transport-aquaplanet} shows the corresponding total meridional OHT (in Petawatts = 10$^{15}$ W) of Case 7 (all OHT active), decomposed into its contributing terms. The magnitude of Ekman transport is driven by the temperature difference between the surface and deep layers while its direction is dictated by the surface wind forcing. Ekman transport (green line) is strongest in the tropics, consistent with Earth-based observations \citep{levitus1987meridional}. This is due to the large temperature contrast between the warm surface water and cooler return flow below, as well as the influence of the Trade winds: via Ekman transport they drive surface waters poleward perpendicular to the wind, inducing equatorial upwelling. However, in the mid-latitudes, the dominant winds are the westerlies, which via Ekman transport drives surface waters equatorward, causing convergence (downwelling) in the subtropics, and a reversal in the meridional OHT. The hemispherical asymmetry in Ekman transport observed in Fig.~\ref{fig:northward-transport-aquaplanet} is likely due to an initial model bias (see Sect.~\ref{subsec:hemisphere-asymmetry} for more details).

For a more direct comparison, we also overlay the corresponding meridional OHT components from \citet{codron2012ekman} in Fig.~\ref{fig:northward-transport-aquaplanet}. The total diffusive transport in the mid-latitudes (crossed grey lines in \citealt{codron2012ekman}; solid blue line in our work) is similar in both studies since the diffusion coefficients were tuned toward the same aquaplanet benchmark \citep{marshall2007mean}.
We further decompose the diffusive transport into its horizontal diffusion (crossed blue) and GM (dashed blue) contributions. Like in \cite{marshall2007mean}, we find that horizontal diffusion plays a negligible role in the meridional heat transport, with a peak less than 0.3 PW at mid-latitudes. This is undoubtedly due to our lower horizontal diffusion coefficient value of 8000 m$^2$s$^{-1}$ compared to \citet{codron2012ekman}. The GM transport, with its relatively high transfer coefficient of 2000 m$^2$s$^{-1}$, peaks around 0.7 PW and brings the global combined diffusive transport (solid blue line) to the same amplitude as that seen in \cite{codron2012ekman}. 
In contrast, Ekman transport in our work (solid green) exhibits larger amplitudes than in \citep[][grey circle-dotted line]{codron2012ekman}, in both the tropics and mid-latitudes, reaching values closer to the Eulerian ones in \citet{marshall2007mean}. This difference primarily arises due to GM transport in the present model. In addition to transporting heat downgradient, GM restratifies the two-layer ocean and strengthens Ekman transport. The poleward shifts of the mid-latitude Ekman trough and diffusive peak relative to \citet{codron2012ekman} likely reflect differences in the atmospheric circulation between the two model versions, which influence surface wind stress and, in turn, OHT profiles. 

With all OHT components accounted for, the general amplitudes resemble those of full AOGCMs (see Figs. 7, 3, 12 and 2 in \citealt{marshall2007mean,brunetti2019co,wu2021coupled,ragon2022robustness}, respectively), despite the much simpler ocean formulation and sparser spatial resolution used here. While the Eulerian (wind-driven) component in \citet{marshall2007mean} reaches values exceeding $\approx 4$ PW, their total OHT peaks only slightly above $\approx 3$ PW. The larger Eulerian peak in fully dynamic models arises because the return flow occurs at greater depths and lower temperatures, increasing the vertical temperature contrast and thereby enhancing the efficiency of Ekman transport. This is then partially compensated by equatorward eddy transport. Our two-layer framework cannot represent this detailed vertical structure; nevertheless, achieving peak transports of 2.25–3.25 PW (solid red line in Fig.~\ref{fig:northward-transport-aquaplanet}) with comparable latitudinal structure indicates that the model captures the first-order behaviour of large-scale OHT. This also represents a clear improvement over the peak transport of $\approx 1.75$ PW reported in \citet[][solid grey line]{codron2012ekman}.


\section{\label{sec:earth}Model validation with modern Earth}

Now that we have better understood the role of the individual OHT regimes using an idealised planet, we apply it on the most widely-studied ocean planet -- modern Earth. The parameters used for our modern Earth simulation are summarised in Table~\ref{tab:physical-parameters}.
The atmosphere is primarily composed of N$_2$, with CO$_2$ at close to modern Earth levels (0.375 mbar $\approx$ 370 ppmv -- value set following \citealt{charnay2013exploring}) and variable H$_2$O. Oxygen and ozone are not included for the sake of simplicity in our radiative transfer calculations. Besides, as pointed out in \cite{charnay2013exploring}, the absence of oxygen only causes small changes in Rayleigh scattering, while that of ozone implies the absence of a stratospheric thermal inversion.

We perform two baseline simulations, one with OHT entirely disabled (Case 1) and another with OHT fully enabled (Case 7). Additionally, we perform a variety of sensitivity tests (see Sect.~\ref{subsec:albedo-gm-sensitivity}). The simulations are initialised with an isothermal atmosphere and surface temperature distribution of 290 K. The simulation without OHT reaches equilibrium within 40 model years, while that with OHT takes around 70 model years to converge. This spin-up time reflects our use of an isothermal 290 K initial state, whereas \citet{codron2012ekman} initialised the simulations from an idealised meridional temperature profile following \citet{neale2000standard}.

Simulating modern Earth also enables model validation through comparison with the extensive suite of observational and reanalysis datasets available for the present-day climate. Specifically, we evaluate our results against state-of-the-art reanalysis data from the National Centers for Environmental Prediction / National Center for Atmospheric Research (NCEP/NCAR; \citealt{kalnay1996}) and the European Centre for Medium-Range Weather Forecasts (ECMWF) European Re-Analysis (ERA-15) dataset \citep{gibson1997era15}. We also compare with observational data from the National Snow and Ice Data Center (NSIDC; \citealt{fetterer-nsidc}) and the National Oceanic and Atmospheric Administration (NOAA) Daily Optimum Interpolation Sea Surface Temperature (DOISST) version 2.1 dataset \citep{huang2021improvements}.


\subsection{\label{subsec:impact-oht-ocean-atmosphere}Impact of ocean heat transport on ocean-atmosphere fluxes}

\begin{figure}[h]
\centering
\includegraphics[width=18.5cm]{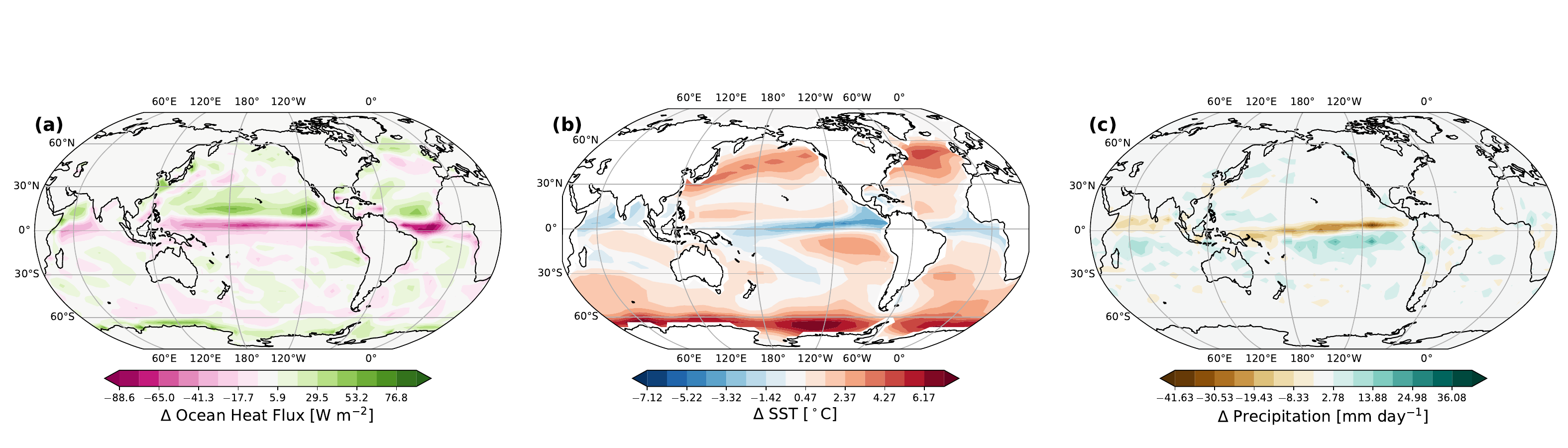}
\vspace{-1em} 
\caption{\label{fig:OHT_Tslab1_Prec_Earth}Example of the causal chain of the coupled ocean-atmosphere system: Impact of ocean heat transport (OHT) on decadal-averaged quantities of our modelled modern Earth. \textbf{(a)} Ocean-atmosphere heat flux from the OHT-on simulation. Note that the OHT-off case has zero ocean heat flux. \textbf{(b)} Corresponding SST difference (OHT-on -- OHT-off). \textbf{(c)} Corresponding precipitation difference.}
\end{figure}

Figure~\ref{fig:OHT_Tslab1_Prec_Earth}a shows the decadal-averaged ocean heat flux from the OHT-on simulation, with net positive values indicating atmospheric heating by the ocean (ocean cooling, upward heat flux) and negative implying atmospheric cooling by the ocean (ocean warming). Alternatively, this plot can be seen as the vertically-integrated convergence of the OHT. We see that our model correctly produces surface ocean cooling (or OHT convergence) in subtropical and high-latitude oceans, and warming in the equatorial region. However, the particularly strong regions of cooling over the Gulf Stream and the Kuroshio currents that are observed on Earth are not captured, since the return flow of the western boundary currents is not explicitly included in our model.

Figure~\ref{fig:OHT_Tslab1_Prec_Earth}b displays the decadal-averaged SST difference between the OHT-on and OHT-off simulations, effectively capturing the thermal response of the surface ocean. There is a clear spatial correspondence between the ocean heat flux (the forcing) and the SST (the response), particularly in the tropics, highlighting the role of OHT in shaping surface temperatures. The colour map is centred around $\Delta T = 0^\circ$C to emphasise regional warming and cooling. 
One of the most prominent features of Fig.~\ref{fig:OHT_Tslab1_Prec_Earth}b is that equatorial SSTs are 5–8$^\circ$C cooler in the OHT-on simulation. As in the aquaplanet case (Fig.~\ref{fig:zonal-temperature-aquaplanet}), this pronounced cold tongue -- especially visible in the eastern Pacific -- is primarily driven by Ekman-induced upwelling. A similar structure is also evident in observations \citep[for e.g.,][]{cromwell1953circulation,wyrtki1966oceanography}.
Our more eastward cold tongue compared to \citet{codron2012ekman} is likely a consequence of Sverdrup dynamics. \citet{codron2012ekman} used a purely frictional Ekman balance at the equator, and noted that the position of the cold tongue is sensitive to the frictional damping coefficient $\epsilon$ (see Sect.~\ref{subsubsec:ekman-transport}), with lower values of $\epsilon$ favouring an equatorially centered, zonally extended cold tongue. In our study, the introduction of a Sverdrup balance near the equator suppresses unrealistically strong cross-equatorial frictional transport (see Sect.~\ref{subsec:hemisphere-asymmetry} for more details), shifting the equatorial circulation into a curl-driven regime and anchoring the upwelling more strongly to the eastern basin.
Meanwhile, mid-latitude SSTs are warmer when OHT is enabled, primarily due to horizontal diffusion, with some contribution from GM advection (also seen in Fig.~\ref{fig:all-scenarios-sst-difference}). The Southern Ocean warms by 7–10$^\circ$C in the OHT-on case, consistent with \citet{codron2012ekman} and \citet{charnay2013exploring}, again driven by GM advection and horizontal diffusion.
Our OHT-on simulation also yields warmer and more realistic tropical SSTs than those in \citet{charnay2013exploring}, with temperatures comparable to NCEP/NCAR reanalysis.

While OHT acts as the forcing and SSTs the response, the downstream consequence is seen in precipitation patterns. Figure~\ref{fig:OHT_Tslab1_Prec_Earth}c shows the corresponding decadal-averaged precipitation difference. As in the aquaplanet simulations, the OHT-off case features a precipitation peak at the equator. However, when OHT is enabled, we get a (weak) double-peaked precipitation structure around the equator due to suppressed evaporation given the Ekman-driven upwelling. Following \citet{codron2012ekman}, we note that lower values of the frictional damping coefficient $\epsilon$ (see Sect.~\ref{subsubsec:ekman-transport}) lead to a more zonally extended equatorial cold tongue.

The annually-averaged position of the ITCZ on Earth lies near 5°N \citep{frierson2013contribution,marshall2014ocean}. This hemispheric preference arises due to a combination of factors, including the unequal distribution of landmasses and the oceanic Meridional Overturning Circulation (MOC). 
The latter plays a particularly important role: \citet{frierson2013contribution} demonstrated that the ITCZ shifts northward in GCM simulations when OHT is included, even in the absence of continents. This suggests that the northward displacement of the tropical rain band is not merely a response to hemispheric radiative asymmetries, but is actively maintained by the MOC, which transports approximately 0.4 PW \citep{frierson2013contribution,marshall2014ocean} of heat across the equator into the Northern Hemisphere, thereby shifting the ITCZ northward.

In our OHT-off simulation, the absence of OHT results in a near-equatorial ITCZ, slightly north-shifted ($\approx$ 3.5°N). This offset is likely a model artefact. Figure~\ref{fig:OHT_Tslab1_Prec_Earth}c shows the precipitation difference between the OHT-on and OHT-off simulations: a negative anomaly in the Northern Hemisphere and a positive one in the Southern Hemisphere indicate a southward migration of the ITCZ when OHT is enabled. In the OHT-on case, the ITCZ is positioned around 3.5°S. This shift arises from: (a) the lack of density-driven MOC in our model, and (b), enhanced Southern Hemisphere warming due to OHT-driven sea ice loss. This Southern Hemisphere warming drives a northward atmospheric energy transport, which shifts the ITCZ to around 3.5°S, as shown in Fig.~\ref{fig:OHT_Tslab1_Prec_Earth}c. This behaviour is consistent with the compensation mechanism discussed in \citet{frierson2013contribution}, where the ITCZ migrates toward the hemisphere receiving more net heating.

\subsection{\label{subsec:impact-oht-meridional-earth}Impact of ocean heat transport on meridional energy redistribution}

\begin{figure}[h]
\centering
\includegraphics[width=10cm]{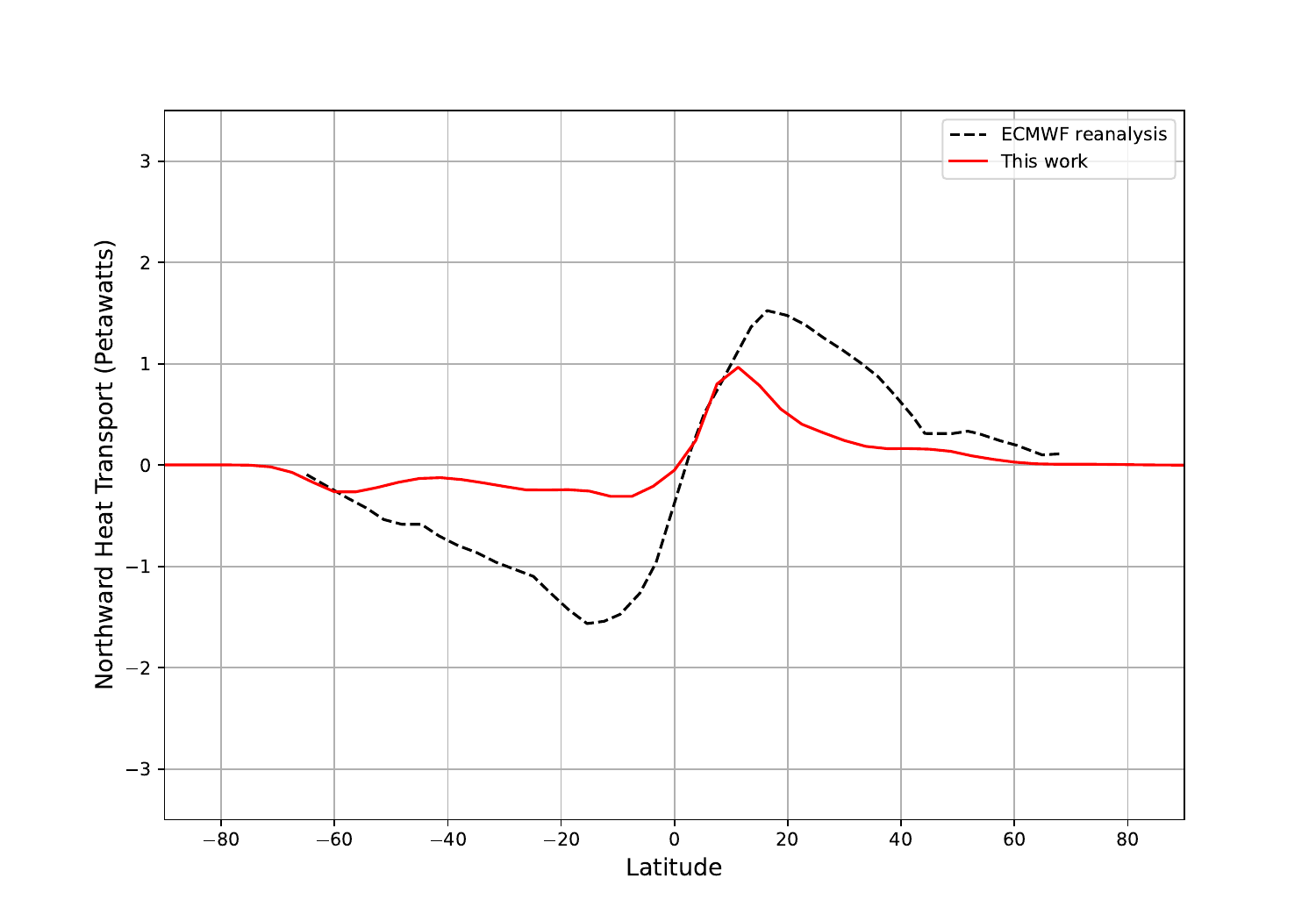}
\vspace{-1em} 
\caption{\label{fig:northward-transport-earth}Total northward meridional ocean heat transport (in Petawatts) for modern Earth, as simulated by our model (solid red line) and derived from ECMWF reanalysis data (black dashed line; \citealt{trenberth2001estimates}).}
\end{figure}

Figure~\ref{fig:northward-transport-earth} presents the meridional ocean heat transport (in petawatts) for modern Earth, comparing results from our model (solid red line) with estimates derived from ECMWF reanalysis (black dashed line; \citealt{trenberth2001estimates}). While our model does not capture the full magnitude of Earth's heat transport, it successfully reproduces the wind-driven component of the circulation well due to the combination of the Ekman and GM parameterisations. This is especially evident in the Northern Hemisphere tropics, where the Pacific Ocean dominates the total transport (see Fig. 5 in \citealt{trenberth2001estimates}).

However, the peak northward OHT in our model ($\approx$ 1 PW) is lower than reanalysis estimates ($\approx$ 1.5 PW), and Southern Hemisphere OHT is also underestimated (--0.3 PW vs. --1.5 PW). These discrepancies stem primarily from the absence of a full-depth MOC in our model. While we include Ekman transport, horizontal diffusion, and the GM parameterisation that together represent key aspects of the wind-driven and eddy-driven components of the MOC, we do not simulate density-driven processes such as deep water formation and diapycnal mixing. These components are critical for reproducing the full strength and structure of the observed MOC, particularly in the Atlantic and Southern Ocean, where density-driven transport dominates (see Fig. 5 in \citealt{trenberth2001estimates}, and also \citealt{frierson2013contribution} and \citealt{marshall2014ocean}).


\subsection{\label{subsec:impact-oht-seasonal-earth}Impact of ocean heat transport on seasonal climate and sea ice}

\begin{figure}[h]
\centering
\includegraphics[width=15cm]{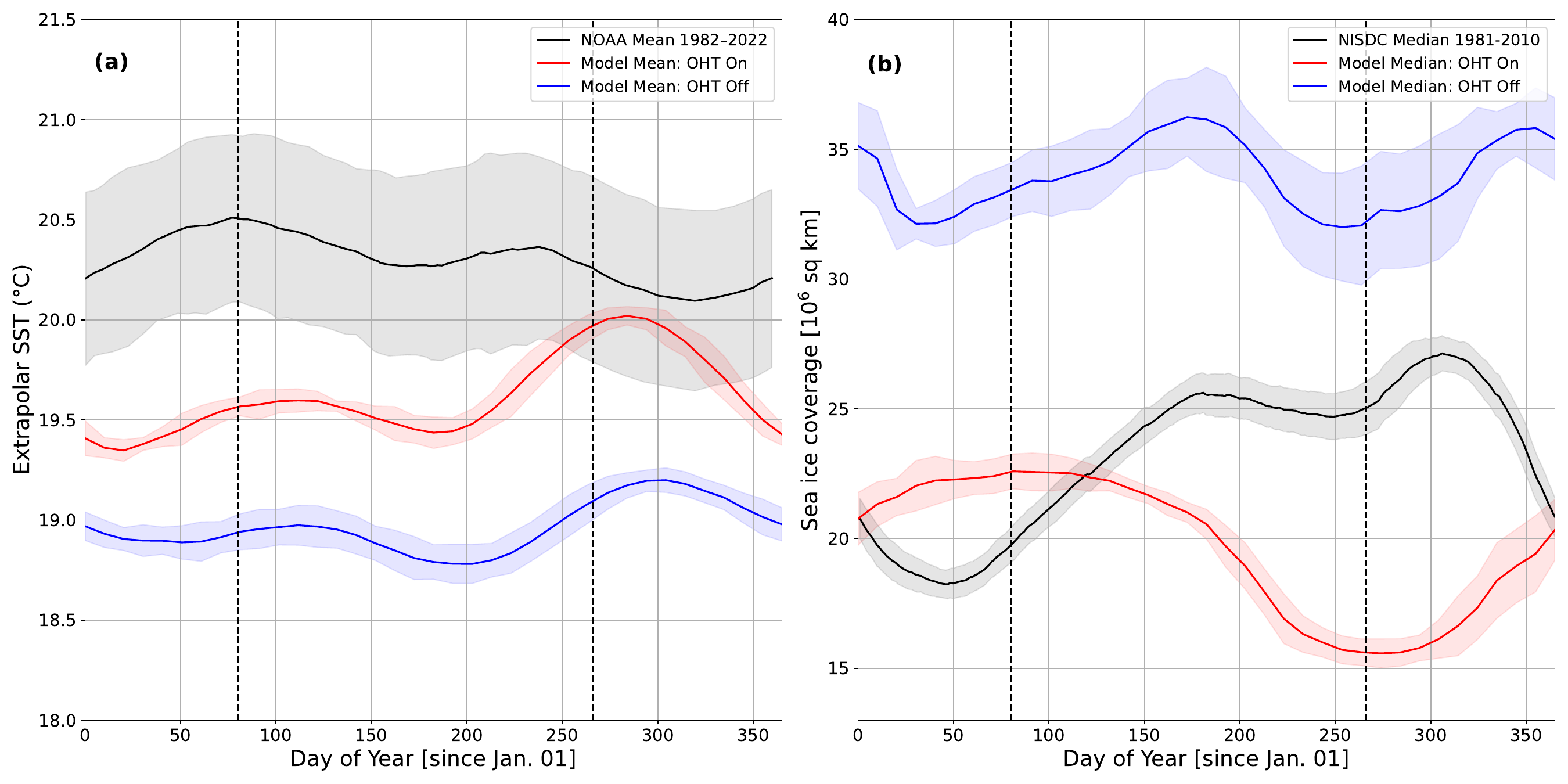}
\vspace{-1em} 
\caption{\label{fig:SST_SeaIce}\textbf{(a)} Seasonal evolution of extrapolar sea surface temperatures (SST; 60°S–60°N), and, \textbf{(b)} global sea ice coverage for NOAA/NSIDC observations (black), for the model with ocean heat transport enabled (red), and disabled (blue). Shaded regions represent the 2$\sigma$ inter-annual variability (30–40 years). Vertical dashed lines indicate the timing of the March and September equinoxes.}
\end{figure}

The seasonal evolution of extrapolar SSTs (60°S–60°N, Fig.~\ref{fig:SST_SeaIce}a) and global sea ice coverage (Fig.~\ref{fig:SST_SeaIce}b) is strongly modulated by the presence or absence of OHT, which affects both the amplitude and phase of their annual cycles. Figure~\ref{fig:SST_SeaIce} compares these diagnostics across three cases: observations (black), the model with OHT enabled (red) and disabled (blue). Shaded regions represent the 2$\sigma$ interannual variability over a 30–40 year period\footnote{The observed SST variability is larger than in the model due to (a) the trend of rising atmospheric CO$_2$ levels between 1982–2022, and (b) multi-year climate variability, for instance, the El Niño–Southern Oscillation (ENSO).}, and vertical dashed lines denote the March and September equinoxes.


Observed SSTs (from NOAA, Fig.~\ref{fig:SST_SeaIce}a) exhibit a characteristic bimodal seasonal pattern, with peaks around the equinoxes, accompanied by corresponding advances and retreats in global sea ice coverage (NSIDC, Fig.~\ref{fig:SST_SeaIce}b), reflecting seasonal polar heating and cooling. The physical origins of the observed SST peaks are rooted in hemispheric asymmetries. The first peak around the March equinox arises from the warming of the vast Southern Hemisphere oceans, whose influence outweighs that of the colder, but smaller Northern Hemisphere ocean area during that time of the year. The second, smaller peak near the September equinox is due to Northern Hemisphere warming, partially offset by cooler Southern Hemisphere SSTs. Our OHT-on and -off simulations reproduce this bimodality to differing extents, but we produce a strong SST peak around the September equinox. This is because our Southern Hemisphere oceans remain too warm due to underestimated Antarctic sea ice (see Fig.~\ref{fig:seaice_north_south}). Additionally, we tend to lag observations by about a month likely due to a slightly elevated thermal inertia in our model ocean and/or the assumption of a fixed mixed-layer depth. The latter overly damps SST variations in regions where the real ocean mixed layer would be shallower -- like in the tropics. We refer the reader to Sects.~\ref{subsec:albedo-gm-sensitivity} and \ref{subsec:model-limitations} for more details.

When OHT is disabled, the climate becomes 1.5–2°C cooler than observed (Fig.~\ref{fig:SST_SeaIce}a), with a muted seasonal SST cycle. Sea ice coverage is unrealistically high year-round -- by approximately 10 million km² (Fig.~\ref{fig:SST_SeaIce}b) -- since the absence of poleward heat transport allows ice to grow in both area and depth. This thick sea ice layer suppresses warming and inhibits ocean-atmosphere coupling, damping SST variability further. Conversely, enabling OHT improves latitudinal heat redistribution, leading to warmer extrapolar SSTs and limiting sea ice extent to values much closer to observations. Although the OHT-on simulation does not fully capture the observed bimodality in sea ice, it more accurately captures the seasonal SST amplitude and the annual averaged values of both SST and sea ice coverage. The annual mean biases are approximately 0.6°C for SST and 3 million km² for sea ice coverage, respectively. The extent of sea ice also influences the planetary bond albedo, which in our OHT-on and -off simulations, is approximately 0.32 and 0.33 respectively, closely matching the value of 0.31 derived from the NCEP/NCAR Reanalysis R-1. These represent a substantial improvement over the value of 0.36 obtained in \citet{charnay2013exploring}, who attributed their high albedo value to an excessive amount of clouds produced at the ITCZ. Our lower and more realistic albedo therefore suggests that we have a reduced ITCZ cloud bias, although a detailed cloud radiative analysis is beyond the scope of this work.


Figure~\ref{fig:SST_SeaIce}b shows that the global median sea ice extent in the OHT-on simulation varies between 16–23 million km² -- much closer to the observed range (18–27 million km²) than the OHT-off case. 
Additionally, in contrast to \citet{charnay2013exploring}, we retain Northern Hemisphere sea ice throughout the year, whereas in their setup it disappeared completely in the summer.
However, both our OHT-on and -off cases still overestimate Northern Hemisphere sea ice (see Fig.~\ref{fig:seaice_north_south}), consistent with findings from \cite{codron2012ekman}. This likely stems from three factors: (a) the absence of a deep overturning circulation transporting heat from the Southern Ocean to the North Atlantic, contributing to both, excess sea ice in the North, and its lack in the South, (b) biases in sea ice albedo (see Sect.~\ref{subsec:albedo-gm-sensitivity}), and (c) the lack of sea ice drift, which promotes excess ice build-up. In contrast, Antarctic sea ice (Fig.~\ref{fig:seaice_north_south}) is underestimated in both model configurations. The same factors (a) and (b) apply here, as well as the missing Antarctic Circumpolar Current (ACC), which, in the real climate system, helps thermally isolate the Antarctic continent and sustain sea ice cover.

We remind the reader that the primary objective of our model is not to exactly reproduce modern Earth’s climate, but to capture its first-order features. This ensures greater confidence in the model’s behaviour when applied to different planetary contexts, whether paleoclimate or exoplanets. Nevertheless, our modern Earth simulations yield surface temperatures that compare well with both observational datasets and previous modelling studies (see Sect.~\ref{subsec:albedo-gm-sensitivity}). With OHT enabled, the simulation produces a global annually averaged surface temperature of 13$^\circ$C, in close agreement with the NCEP/NCAR reanalysis temperature of 14.0$^\circ$C for the 1981–2010 period. In contrast, the simulation without OHT yields a cooler average temperature of 12$^\circ$C.
\citet{charnay2013exploring}, by comparison, obtained a warmer OHT-enabled mean temperature of 14.9$^\circ$C. They note that this likely results from a combination of an overly strong meridional heat transport and an enhanced greenhouse effect due to a larger amount of high clouds, as well as the seasonal disappearance of Arctic sea ice in summer. In our work, Arctic sea ice is retained throughout the year, which contributes to a slightly cooler climate.

Beyond global average temperatures, the inclusion of OHT also improves other key diagnostics, such as extrapolar SSTs and total sea ice extent, bringing them into closer alignment with observations. These quantities play a central role in the planet’s energy balance—through their influence on albedo and outgoing longwave radiation—and are especially relevant for exoplanet climate modelling. Given the limited constraints on exoplanet climates, such agreement with Earth benchmarks is a promising validation of our model’s physical realism. 

\section{\label{sec:discussion}Discussion}
Given the successful validation of our model against aquaplanet and modern Earth benchmarks, this section explores the broader implications and insights of our dynamical slab ocean framework. 

\subsection{\label{subsec:gm-discussions}The role of the Gent-McWilliams scheme: Influence on vertical structure and ocean heat transport}

\begin{figure}[h]
\centering
\includegraphics[width=15cm]{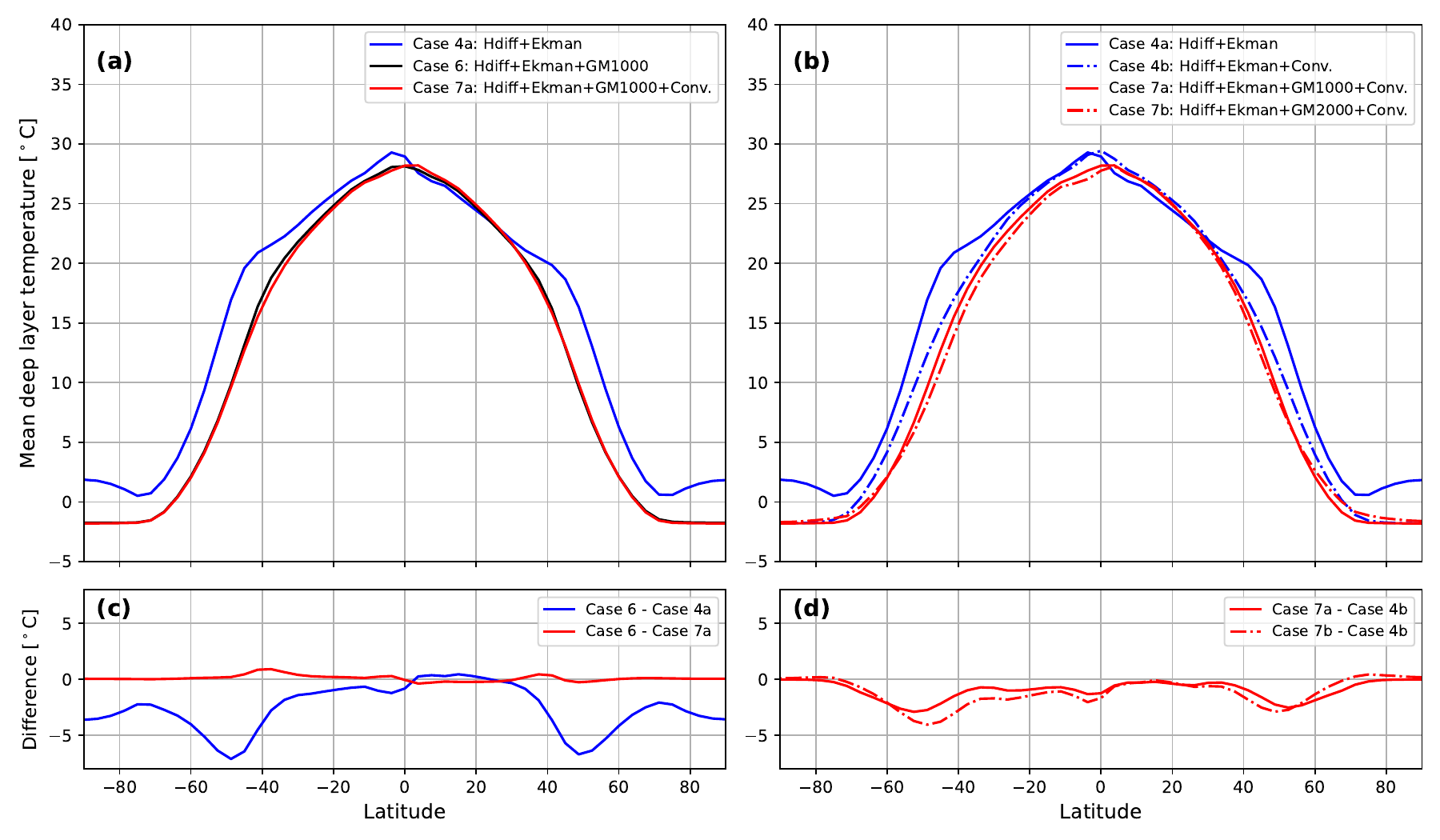}
\vspace{-1em} 
\caption{\label{fig:deep-layer-temperature-Dec2025}Gent-McWilliams (GM) restratification relative to convective adjustment: Zonally averaged deep ocean layer temperatures in the aquaplanet simulations. \textbf{Top panels:} Mean deep layer temperature for the selected simulations; \textbf{Bottom panels:} Differences between cases.}
\end{figure}

One of the additions we introduce to the Generic-PCM's dynamical slab ocean model is the Gent-McWilliams (GM) parameterisation \citep{gent1990isopycnal}. The GM scheme enables isopycnal mixing and modifies the large-scale ocean temperature structure (Fig.~\ref{fig:deep-layer-temperature-Dec2025}) through horizontal and vertical heat redistribution. In our two-layer model, this redistribution alters the temperature contrast between the surface and deep layers that drives Ekman transport (see Sect.~\ref{subsubsec:ekman-transport}), thereby introducing a coupling between eddy-induced and wind-driven circulation.

To assess the relative importance of GM and convective adjustment, we compare a hierarchy of aquaplanet simulations that systematically include or exclude these processes. Figure~\ref{fig:deep-layer-temperature-Dec2025}a shows the decadal-averaged zonal deep layer temperatures of the relevant simulations. Case 6 (horizontal diffusion + Ekman + GM; solid black line) exhibits substantially cooler deep layer temperatures at mid- to high latitudes than the simulation without GM (Case 4a; horizontal diffusion + Ekman; solid blue line), with differences reaching 5–7$^\circ$C (see Fig.~\ref{fig:deep-layer-temperature-Dec2025}c). In contrast, adding convective adjustment on top of GM (Case 7a; solid red line) produces only negligible additional changes (Fig.~\ref{fig:deep-layer-temperature-Dec2025}c). This indicates that, under the conditions tested here, the inclusion of GM substantially modifies vertical stratification such that the contribution of convective adjustment to deep layer temperatures is minor.

We next consider Case 4b (horizontal diffusion + Ekman + convective adjustment; dash-dotted blue line). Compared to Case 4a, convective adjustment reduces the deep layer temperatures (Fig.~\ref{fig:deep-layer-temperature-Dec2025}b), but this cooling is weaker than that produced by GM only (Fig.~\ref{fig:deep-layer-temperature-Dec2025}c). Still, this highlights that the effects of GM and convective adjustment are not simply additive. To further isolate the role of GM when convective adjustment is already active, we compare Case 4b with Case 7a. This reveals that the inclusion of GM, which drives eddy-induced heat redistribution, produces additional mid-latitude cooling of 2–3$^\circ$C (Fig.~\ref{fig:deep-layer-temperature-Dec2025}d). This shows that GM has a distinct influence on the deep ocean even in the presence of convective adjustment.

To test sensitivity to the strength of the GM scheme, we performed Case 7b, in which the GM transfer coefficient is increased to 2000~m$^2$~s$^{-1}$ (compared to 1000~m$^2$~s$^{-1}$ in Case 7a). Due to the increased coefficient, mid-latitude cooling reaches 3-4$^\circ$C (Fig.~\ref{fig:deep-layer-temperature-Dec2025}d), consistent with stronger eddy-induced redistribution of heat. This could be particularly relevant for slowly rotating planets, for which mixing coefficients could be substantially larger \citep[e.g.,][]{cullum2014importance}. This will be investigated in future work.

Overall, these results show that convective adjustment alone does not reproduce the equilibrium climate obtained when GM is included. Even when convection is active, adding GM leads to systematic cooling of the deep ocean at mid-latitudes, whereas adding convective adjustment to a configuration that already includes GM produces only minor additional changes.

This is further emphasised in another set of simulations to assess whether an enhanced horizontal diffusion and convective adjustment can substitute for GM. For this, we compared a \citet{codron2012ekman}-style configuration (horizontal diffusion coefficient of 25000~m$^2$~s$^{-1}$, Ekman transport and convective adjustment) with the recommended configuration of the present model including GM (Case 7b). Our findings show that even with substantially enhanced horizontal diffusion and active convection, the absence of GM leads to systematically warmer mid-latitude deep-layer temperatures. This comparison is discussed in detail in Appendix~\ref{subsec:vertical-ocean-structure}.

\begin{figure}[h]
\centering
\includegraphics[width=18cm]{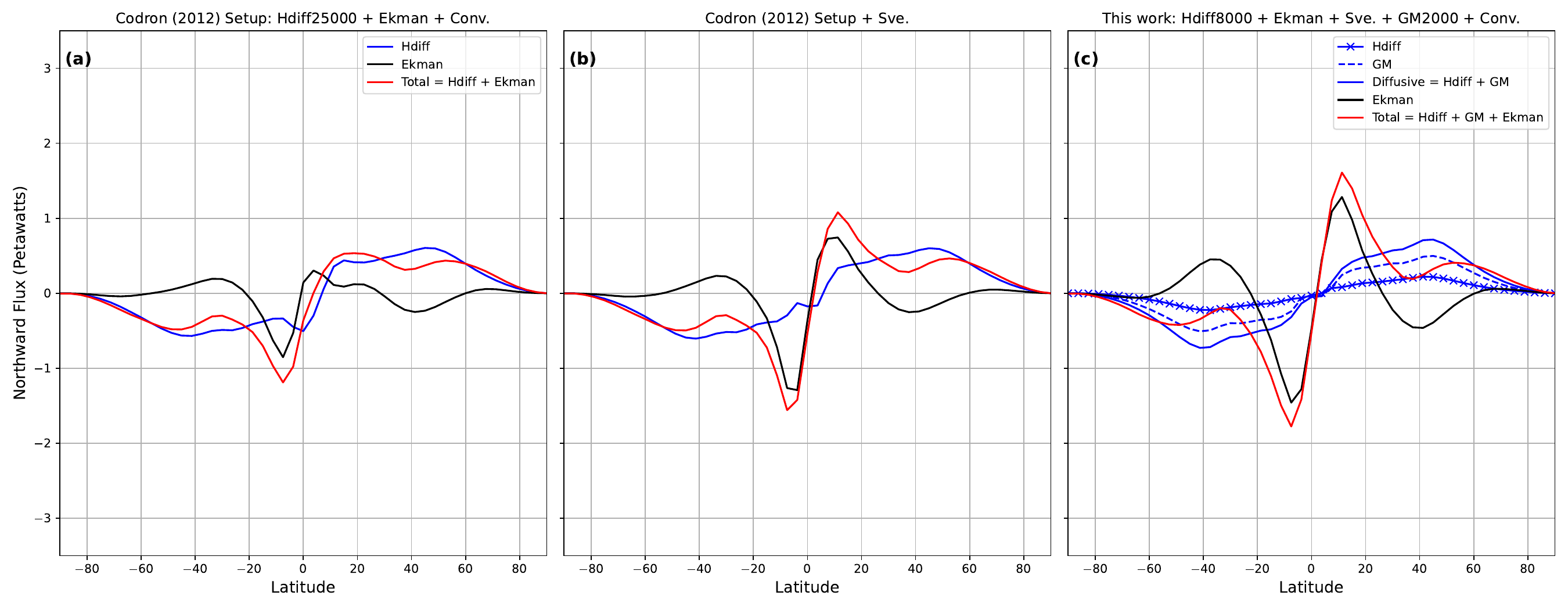}
\vspace{-1em} 
\caption{\label{fig:OHT-profiles-comparison}Total and decomposed ocean heat transport (OHT) for \textbf{(a)} a Codron-style configuration (high horizontal diffusion, Ekman transport without Sverdrup balance and convective adjustment), \textbf{(b)} the same configuration with the Sverdrup balance enabled, and \textbf{(c)} the recommended configuration of the present model (reduced horizontal diffusion, Gent–McWilliams advection, Ekman transport with Sverdrup balance and convective adjustment).}
\end{figure}

To further contextualise the present model, we compare the OHT structure obtained using the configuration of \citet{codron2012ekman}. Figure~\ref{fig:OHT-profiles-comparison} presents total and decomposed meridional OHT profiles for three configurations, allowing us to isolate the effects of the Sverdrup balance, GM transport and updated parameter choices.

Panel~(a) reproduces the \citet{codron2012ekman} oceanic configuration as closely as possible, with horizontal diffusion (coefficient of 25000~m$^2$~s$^{-1}$), Ekman transport (without the Sverdrup balance) and convective adjustment. In this configuration, the OHT exhibits a pronounced hemispheric asymmetry, discussed in detail in Sect.~\ref{subsec:hemisphere-asymmetry}. Interestingly, the OHT profile in \citet{codron2012ekman} (see Fig.~\ref{fig:northward-transport-aquaplanet}) is quite symmetric, despite the absence of Sverdrup balance in that study. This could point towards a decreased sensitivity of the atmospheric model to surface temperature variations in that version of the GCM. Correspondingly, it implies that the current atmospheric model is either sufficiently or overly sensitive to surface temperature variations. While investigating the related sensitivity is beyond the scope of this work, its physical reasoning is explained in Sect.~\ref{subsec:hemisphere-asymmetry}.

Panel~(b) shows the same Codron-style configuration but with the Sverdrup balance enabled. This largely restores hemispheric symmetry in the OHT, showing its negative-feedback-induced stabilising role (see Sect.~\ref{subsec:hemisphere-asymmetry}). However, despite this improvement, the amplitude of tropical Ekman transport remains suppressed relative to AOGCM benchmarks \citep[e.g.,][]{marshall2007mean,ragon2022robustness}. This indicates that the Sverdrup balance alone is insufficient to recover a realistic OHT structure when GM transport is absent.

Panel~(c) presents the recommended oceanic configuration of the present model, which uses a reduced horizontal diffusion coefficient (8000~m$^2$~s$^{-1}$), GM transport (2000~m$^2$~s$^{-1}$), Ekman transport with the Sverdrup balance, and convective adjustment. Here, the tropical OHT peak -- primarily driven by Ekman transport -- reaches amplitudes closer to those in fully coupled AOGCM simulations (e.g., Fig.~7 of \citealt{marshall2007mean}; hot-state solutions in Fig.~2 of \citet{ragon2022robustness}) than those obtained either in \citet[][Fig. 4]{codron2012ekman} or in the \citet{codron2012ekman} setups with the current model. This increase arises since GM transport modifies the temperature contrast between the surface and deep layers, thereby affecting the strength of Ekman transport (Eq.~\ref{eqn:ekman-equation}). Likewise, the Ekman-driven mid-latitude trough is more pronounced due to GM transport in panel~(c), resembling the amplitude of the Eulerian transport component in AOGCM simulations (Marshall et al., 2007; Fig. 7). Compared to the $\approx 30^\circ$ peak in Fig.~4 of \citet{codron2012ekman}, the latitude of the diffusive transport peak is also shifted poleward to approximately 45$^\circ$, in closer agreement with \citet{marshall2007mean}. Its peak amplitude (around 0.6 PW) is provided solely by horizontal diffusion in panels (a) and (b). In contrast, the present model decomposes diffusive transport into its horizontal diffusion and GM transport components, yielding a slightly larger combined diffusive peak of $\approx 0.75$ PW in panel~(c). 

The tropical OHT peak is also sensitive to the system’s climatic attractor. In panel~(c), which corresponds to a “hot state” \citep[see][]{brunetti2019co} with no sea ice, the tropical OHT peak is approximately 1.75 PW, consistent with the hot state solution of \citet[][Fig. 2]{ragon2022robustness} obtained with the MITgcm. In contrast, in simulations tuned to reproduce an Earth-like sea-ice distribution (e.g., Fig.~\ref{fig:northward-transport-aquaplanet} in this work), we obtained peak transports of 2–3 PW, similar to the warm state attractor in \citet{ragon2022robustness}.

These comparisons demonstrate that the combined inclusion of the novel features of the model: Sverdrup balance, GM transport, and updated parameter choices, leads to an improvement over \citet{codron2012ekman}. The resulting OHT structure more closely resembles that of fully dynamic AOGCMs, despite the relative simplicity of our model.

\subsection{\label{subsec:hemisphere-asymmetry}The role of the Sverdrup balance scheme: Suppression of hemispheric asymmetries and numerical instabilities}

Previous aquaplanet simulations with obliquity using the Generic-PCM (e.g., \citealt{chaverot2023first}) revealed a persistent pronounced hemispheric asymmetry in surface temperature and cloud cover, even after year-long averaging. In \cite{chaverot2023first}, simulations were initialised starting from the Vernal Equinox and without OHT. The resulting asymmetry was attributed to an initial hemispheric bias that triggered a net north–south atmospheric heat transport, allowing one Hadley cell to dominate over the other. This induced a feedback that persisted throughout the simulation.

\begin{figure}[h]
\centering
\includegraphics[width=10cm]{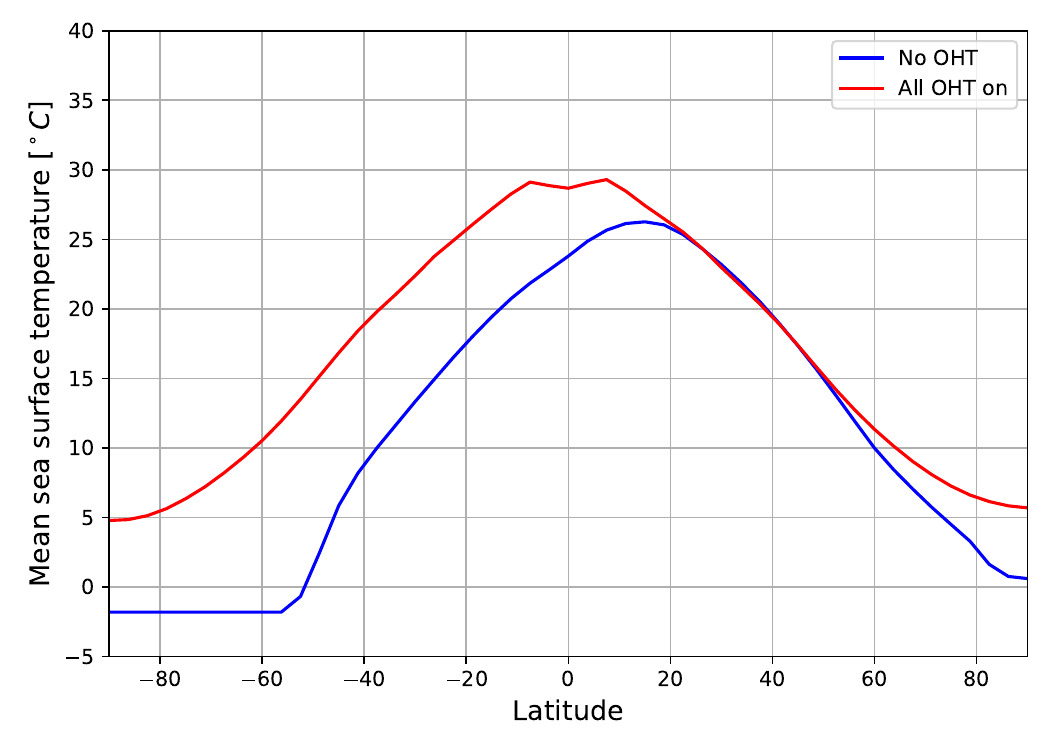}
\vspace{-1em} 
\caption{\label{fig:hemispheric-asymmetry}Ocean heat transport (OHT) suppresses hemispheric climate asymmetries caused by initialisation biases: zonally averaged sea surface temperatures for our oblique aquaplanet setup, with OHT disabled (blue) and enabled (red).}
\end{figure}

We propose an additional explanation involving the role of OHT. Our oblique aquaplanet simulations show that when OHT is disabled (blue line in Fig.~\ref{fig:hemispheric-asymmetry}), the hemispheric bias imprinted at model initialisation is not corrected by the system. In the absence of OHT, there is no mechanism—apart from atmospheric transport—to redistribute excess energy across hemispheres, triggering a cascade of feedbacks. The initially warmer Northern Hemisphere delays sea ice growth, while the Southern Hemisphere cools rapidly due to atmospheric transport acting on short timescales. This cooling is amplified by the ice–albedo feedback, leading to early and extensive sea ice formation, which locks in a strong hemispheric asymmetry. Atmospheric heat redistribution alone is insufficient at evening out hemispherical differences -- and can even amplify them -- through feedbacks involving the cross-equatorial Hadley cell, which develops with ascending motion in the warmer hemisphere. The associated cloud cover and moisture content further modulate the radiation budget, reinforcing the asymmetry. 

When OHT is on (red line in Fig.~\ref{fig:hemispheric-asymmetry}), the ocean helps smooth out temperature gradients, mitigating the imprint of the initial bias. At model start, heat stored in the ocean is transported from lower to higher latitudes in both hemispheres through the mechanisms discussed in Sect.~\ref{subsec:heat-transport}. This redistribution limits excessive sea ice formation in the South and prevents the North from staying anomalously warm. The ocean circulation forced by the cross-equatorial Hadley cell will also transport heat across the equator towards the cold hemisphere, with opposite feedbacks from the ones associated with atmospheric transport. In general, strong persistent hemispheric asymmetries in aquaplanet setups are unphysical and typically reflect initial condition biases and/or numerical artefacts. The key takeaway is that \textit{both} oceanic and atmospheric heat transport are essential for suppressing such asymmetries and maintaining a physically realistic climate state.

In our model, this stabilising effect of OHT is largely driven by the wind-curl-sensitive Sverdrup balance scheme implemented near the equator (Sect.~\ref{subsubsec:ekman-transport}). To assess the impact of this scheme, we conducted two simulations with Sverdrup transport disabled, one initialised from the converged Case 7 (non-oblique) state, and the other from an isothermal 290 K start to check for dependence on the start state.

\begin{figure}[h]
\centering
\includegraphics[width=15cm]{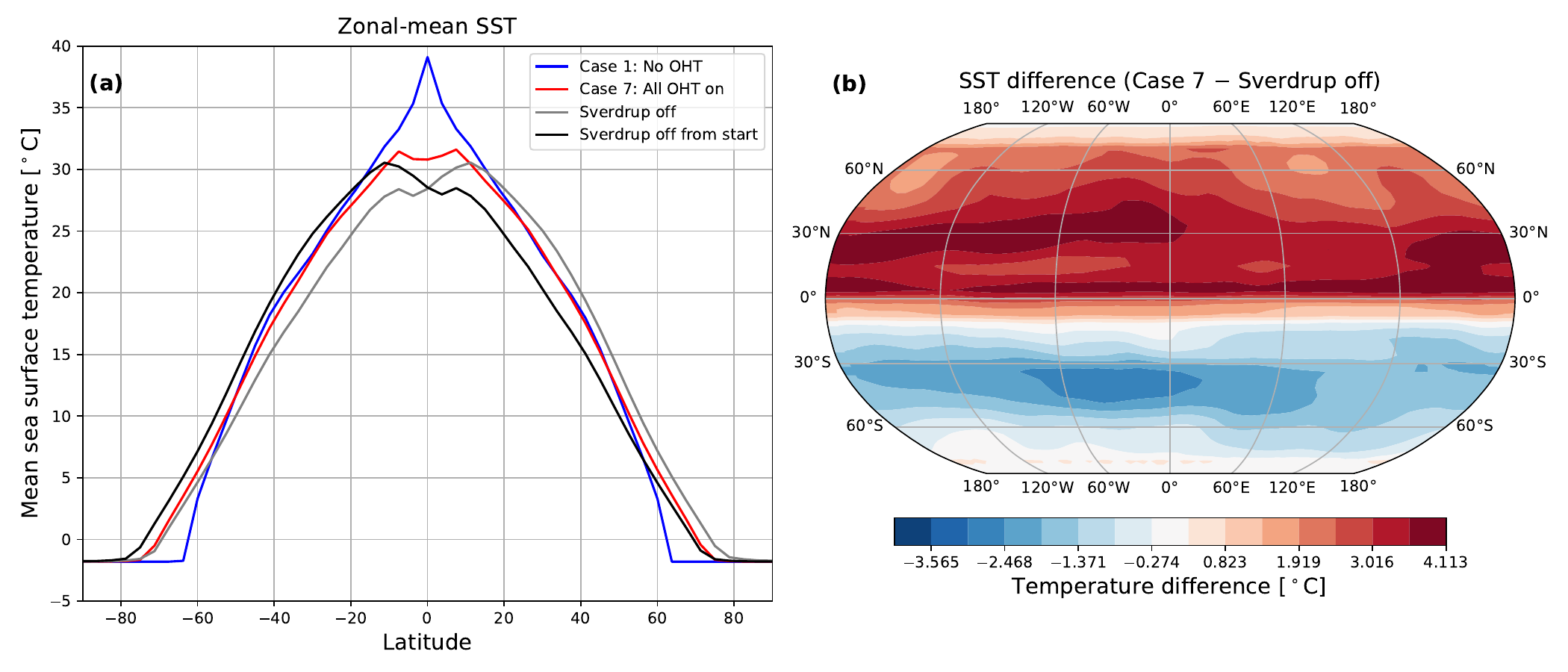}
\vspace{-1em} 
\caption{\label{fig:sverdrup-zonal-map}The stabilising influence of the Sverdrup balance on the climate: 
\textbf{(a)} Decadal-mean zonally averaged sea surface temperature (SST) for simulations with the Sverdrup balance enabled (Case 7) and disabled. Two Sverdrup-off simulations are shown, one initialised from the converged Case 7 state and (ii) an isothermal 290 K state.
\textbf{(b)} Decadal-mean SST difference (Case 7 -- Sverdrup-off).}
\end{figure}

Without the Sverdrup balance, near-equatorial OHT responds purely through frictional transport (see Eq.~\ref{eqn:massflux}). In this case, the cross-equatorial OHT occurs in the same direction as the surface wind. This generates an SST asymmetry (Fig.~\ref{fig:sverdrup-zonal-map}a) that shifts the ascending branch of the Hadley cell, thereby amplifying the cross-equatorial wind anomaly (Fig.~\ref{fig:sverdrup-winds}a). This leads to a strong hemisphere-wide dichotomy of temperature anomalies (Fig.~\ref{fig:sverdrup-zonal-map}b). Moreover, our two Sverdrup-off experiments converge to opposite asymmetric states, illustrating this bistability (Fig.~\ref{fig:sverdrup-zonal-map}a).

With the Sverdrup balance enabled (Case 7), the near-equatorial transport responds to the wind stress curl rather than its frictional meridional component. For hemispherically symmetric forcing, the wind stress curl remains symmetric about the equator. Importantly, the resulting cross-equatorial OHT occurs opposite to the surface wind direction, introducing a negative feedback that suppresses growing inter-hemispheric asymmetries. As a result, the coupled system converges to a more symmetric equilibrium, both for SST (Fig.~\ref{fig:sverdrup-zonal-map}a) and net surface winds (Fig.~\ref{fig:sverdrup-winds}b). Consequently, the inclusion of the Sverdrup balance in our model plays a crucial role in preventing the locking-in of hemispheric biases and maintaining a physically consistent climate.

With further experiments, we found that enabling OHT also enhances numerical stability in the Generic-PCM. In the OHT-off configuration, the model frequently encountered dynamical instabilities during long integrations. These usually manifest as spurious temperature dips near the surface or at the top of the atmosphere, often triggering radiative transfer scheme failures (temperatures dropping below the lower bounds of the lookup tables), causing the model to crash. These issues are largely absent when OHT is enabled. A likely explanation is that OHT efficiently smooths meridional temperature gradients -- particularly at smaller scales due to the diffusive components -- thereby reducing the likelihood of instabilities and excessive vertical motions. These OHT-smoothened temperature fields may also improve numerical stability by reducing sharp vertical gradients that can otherwise trigger Courant–Friedrichs–Lewy (CFL) violations in the model’s dynamical core.

\subsection{\label{subsec:q-flux}Comparison with a q-flux ocean}

One approach in low-complexity ocean modelling is the q-flux method, where OHT is prescribed as a fixed meridional heat flux (q-flux) based on observational data or AOGCM simulations. While this avoids the computational cost of explicitly solving ocean dynamics, it requires an initial, expensive reference simulation to determine the appropriate OHT. In the context of exoplanets, where observational constraints are limited and reference dynamical ocean simulations are typically unavailable \citep[with some exceptions: for e.g.,][]{del2019habitable,yang2020transition,batra2024climatic}, a physically motivated q-flux cannot be prescribed. More critically, if an OHT is imposed rather than emergent, it cannot dynamically respond to changes in the atmosphere, ocean, or external forcing. As a result, key climate feedbacks -- such as adjustments in meridional heat transport due to sea ice growth, atmospheric circulation shifts, or changes in composition (e.g., chemistry, greenhouse gases) -- are not captured self-consistently. 

In contrast, our dynamical slab ocean model allows for an emergent OHT based on ocean-atmosphere interactions. This is particularly important in cases where strong feedbacks develop -- such as the ice-albedo effect in our simulations or the water vapour feedback in \cite{chaverot2023first} -- and more broadly because atmospheric and oceanic heat transports are strongly coupled via surface winds.
We note, however, that the present model is primarily designed for Earth-like rotation rates and parameter regimes. For instance, our Sverdrup balance may overestimate transport for slowly rotating planets (Sect.~\ref{subsubsec:ekman-transport}), and diffusion coefficients are calibrated using Earth-rotation aquaplanet benchmarks (see Sects.~\ref{subsubsec:gent-mcwilliams} and \ref{subsec:model-limitations}). Given this, our approach currently shares with q-flux methods a reliance on well-characterised regimes. Nevertheless, a key distinction is that even within these regimes, OHT in our model is not prescribed but responds interactively to changes in atmospheric forcing and climate state. Extending the framework to other planetary contexts is a natural direction for future work. 

\subsection{\label{subsec:albedo-gm-sensitivity}Model sensitivity to GM and albedo coefficients}
\label{sec:tests}

\begin{figure}[h]
\centering
\includegraphics[width=15cm]{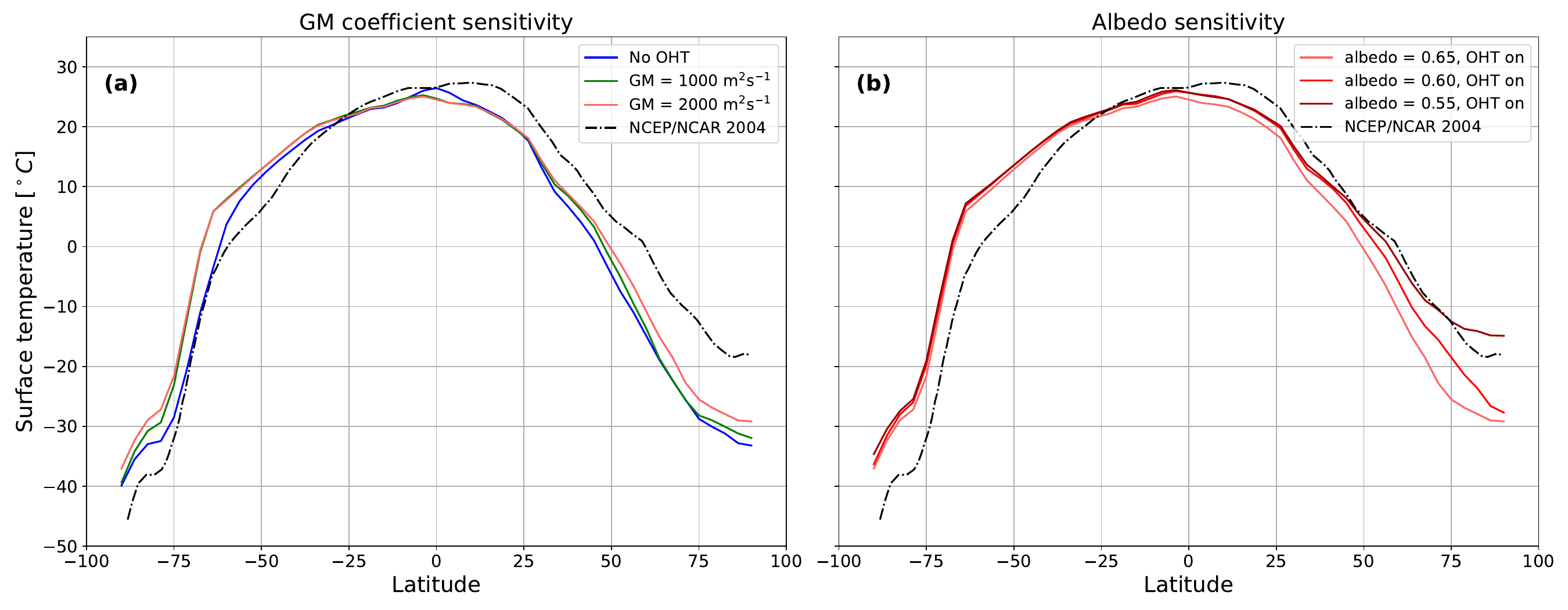}
\vspace{-1em} 
\caption{\label{fig:earth-zon-avg} Zonally averaged surface air  temperature for modern Earth simulations under two different sensitivity experiments. \textbf{(a)} Gent-McWilliams (GM) coefficient sensitivity test comparing model simulations with two different GM values against NCEP/NCAR reanalysis (black dashed line). \textbf{(b)} Albedo sensitivity test comparing model simulations with three different sea ice/snow albedo values against reanalysis data.}
\end{figure}

Figure~\ref{fig:earth-zon-avg}a explores model sensitivity to the GM and horizontal diffusion coefficients. In the OHT-off simulation (blue), surface temperatures at high northern latitudes (60°-90°N) are up to 15°C colder than the NCEP/NCAR reanalysis (black dashed line). Introducing OHT by increasing the GM coefficient (green and red lines, for 1000 and 2000 m$^2$s$^{-1}$, respectively) and/or varying the horizontal diffusion (8000 to 25000 m$^2$s$^{-1}$) warms the Arctic modestly, raising temperatures by about 3-4°C. However, this is insufficient to eliminate the persistent cold bias. Seasonal analysis reveals that while summer Arctic temperatures fall within expected bounds (-10°C to 0°C), average winter temperatures plunge to $\approx$ -50°C -- approximately 20°C colder than observed. This suggests that a more dominant process is responsible, likely involving a combination of atmospheric transport inefficiencies, radiative imbalances, and/or model tuning choices such as surface albedo.

Figure~\ref{fig:earth-zon-avg}b illustrates the model's sensitivity to sea ice and snow albedo, which is reduced from 0.65 to 0.60 and 0.55 for the OHT-on simulation. Lowering the albedo does mitigate the Arctic cold bias, with a stronger warming effect than seen in the GM experiments. However, this comes at the cost of Southern Hemisphere sea ice almost completely disappearing, while concurrently reducing annual global sea ice extents to unrealistic levels (10 million sq. km), introducing a new discrepancy. The systematic Southern Hemisphere warm bias also hints at a hemispheric imbalance, where the (orbital-parameter-induced) excess heat in the south is not being effectively redistributed to the Northern Hemisphere, despite the heat transport provided by the ocean. Additionally, the atmospheric model of the GCM may under-represent poleward moist static energy transport, particularly via latent heat. Together, these factors suggest a complex interplay between radiative, oceanic, and atmospheric processes that warrants deeper investigation. Nevertheless, we reiterate that these discrepancies remain within acceptable bounds, given that our model is not built with the intention of precisely reproducing modern Earth's climate but rather that of exoplanets, where observational constraints are far more uncertain.

\subsection{\label{subsec:advances-over-previous-models}Advances over previous dynamical slab ocean implementations}

We summarise below the key developments of the present model relative to earlier implementations \citep{codron2012ekman,charnay2013exploring} that enhance its suitability for exoplanet and paleoclimate applications.

Two of our improvements are particularly relevant for simulating M-star planets, which are of high interest given their abundance in the galaxy and their observational accessibility. First, the inclusion of a spectrally dependent sea ice and snow albedo (Sect.~\ref{subsec:sea-ice-evolution}) enables more realistic radiative feedbacks on planets receiving non-solar spectra. For M-dwarfs, broadband albedos of snow and ice can be 25–55\% lower than for G-type stars \citep{joshi2012suppression}, making this refinement essential for accurate climate modelling. This is also critically important to better interpret observational data from the JWST and in the future, the ELT. 

Second, the introduction of GM transport represents more than a numerical improvement. Through its transfer coefficient, the GM scheme acts as a tuning parameter that controls the effective scale of mesoscale eddies (see Sect.~\ref{subsec:model-limitations}), enabling exploration of different rotational regimes, notably slow rotators such as TRAPPIST-1e (rotation period 6.1 days) and Proxima b (11.2 days), for which mixing is expected to differ substantially from Earth \citep[e.g.,][]{cullum2014importance}. In earlier models lacking GM \citep{codron2012ekman,charnay2013exploring}, no consistent framework existed to account for such regime-dependent mixing.

Another strength of the present framework is its flexible architecture. Individual oceanic processes can be selectively (de)activated through configuration flags, allowing users to isolate processes, depending on their need. Moreover, the modular design enables technically straightforward integration of additional processes, such as a simplified sea-ice drift scheme (see Sect.~\ref{subsec:model-limitations}), also important for exoplanet applications \citep{yang2020transition}.

Finally, the present ocean model is fully parallelised, in contrast to the implementations of \citet{codron2012ekman} and \citet{charnay2013exploring}. This enables efficient long integrations and large ensemble simulations (see Appendix~\ref{sec:computation-speed} for more details), which are essential for exoplanet and paleoclimate studies where observational constraints are sparse.

\subsection{\label{subsec:model-limitations}Caveats and opportunities for future work}

One of the driving applications for developing the dynamical slab ocean model is to simulate the climates of terrestrial exoplanets. Given that the model is flexible, relatively easy modifications can be made. For example, some exoplanets are expected to experience non-negligible geothermal heat fluxes -- this additional energy source can be incorporated in the model as a constant or spatiotemporally varying ground-heating term (see Eq.~\ref{eqn:general-equation}). 
Internal tidal heating may also represent an additional and potentially dominant heat source for some close-in terrestrial exoplanets \citep[e.g.,][]{driscoll2015tidal,dobos2019tidal,bolmont2026tidal}. Furthermore, tidal forcing of the ocean can influence the large-scale circulation and climate, particularly for aquaplanets orbiting low-mass stars, where strong tides can significantly influence ocean mixing and heat transport \citep[e.g.,][]{si2022planetary,di2025nonlinear}. These additional physical processes represent promising future directions.

Moreover, many of the currently known terrestrial exoplanets are expected to be in synchronous rotation, particularly those orbiting close to M-dwarf stars. In such cases, the planet's rotation period equals its orbital period -- typically ranging from 5 to 15 Earth days for temperate planets. This can place these planets in dynamical regimes distinct from Earth-like rotators \citep[for e.g., see][]{sergeev2022trappist,sergeev2022bistability}. Most climate model parameters, including those governing subgrid-scale mixing, are traditionally calibrated for Earth-like rotation rates. For instance, the diffusion coefficients discussed in Sect.~\ref{subsec:heat-transport} were selected to match the meridional ocean heat transport of an Earth-like aquaplanet \citep{marshall2007mean}. However, the reduced Coriolis parameter \( f \) on slowly rotating planets will lead to a larger Rossby deformation radius (since \( L_D \sim \sqrt{g'H}/f \), where $g'$ and $H$ are the reduced gravity and scale height respectively), which sets the typical scale of baroclinic eddies, from 10 to 100 km on Earth's ocean. This suggests that eddy-driven mixing processes may differ substantially on such climates (see \citealt{showman2010atmospheric} and \citealt{cullum2014importance}). Though our model does not include the density-driven component of the MOC, the GM scheme modifies its temperature-driven part, and leads to high-latitude surface warming and low-latitude depth cooling. This is qualitatively similar to the behaviour reported by \citet{cullum2014importance}, who showed that slower planetary rotation rates enhanced MOC strength and led to comparable thermal redistribution. This is a strong motivation for us to tune the GM coefficient in future exoplanet implementations of the model.

A related limitation arises from our assumption of a globally uniform, fixed mixed-layer depth (MLD). In reality, the MLD varies with location and season due to wind forcing, ocean currents, and surface heating \citep[e.g.,][]{mccreary2001influences,kara2003mixed}. \citet{olson2020oceanographic} also showed that mixed layer seasonality increases with planetary obliquity. The fixed MLD approach simplifies calculations of ocean heat transport, but can introduce biases. First, regions that should have a deeper MLD, such as high-latitude oceans during winter, may exhibit larger or faster surface temperature variations due to an artificially small heat capacity. Conversely, regions that should have a shallower MLD, such as tropical oceans, may experience overly dampened temperature variations. This limitation could impact the accuracy of (short-timescale) climate feedbacks, particularly the ice-albedo feedback, since a dynamically varying MLD would allow for more realistic oceanic heat uptake and redistribution. Choosing a representative MLD is particularly challenging for exoplanets, where no observational constraints exist -- akin to the challenge of prescribing a q-flux. A promising direction for future work is to adopt an “Ekman mixed-layer ocean” scheme, which uses multiple slab layers and includes explicit vertical mixing. This approach, demonstrated by \citet{hsu2022hierarchy} -- who combined it with the Ekman transport parameterisation of \citet{codron2012ekman} -- may offer a more physically robust framework than using a fixed-depth single-layer ocean.

Our model includes salinity solely as a modifier of the freezing point of seawater. This is an important modification as a depressed freezing point inhibits sea ice formation -- leading to reduced global sea ice cover with increasing salinity. \citet{olson2022effect} demonstrated that higher salinities tend to warm planetary climates -- offering a potential solution to the Faint Young Sun Paradox for Archean Earth -- and attributed this warming primarily to changes in density-driven ocean dynamics rather than to freezing point depression. Building on this, \citet{batra2024climatic} found that while salinity has a strong, nonlinear impact on G-star planets, leading to abrupt sea ice loss and surface warming, its effect on M-dwarf planets is more gradual, with minimal associated warming.  
Our model does not represent the dynamical effects of salinity -- such as its role in setting ocean density or driving thermohaline circulation. As highlighted by \citet{codron2012ekman}, this can lead to an underestimation of convective exchanges in high-latitude regions and inadequate cooling of the deep ocean. The absence of salinity also prevents the model from capturing salt rejection during sea ice formation, a process that increases local water density and promotes vertical mixing.
While density-driven circulation due to salinity is important in shaping OHT \citep[e.g.,][]{cullum2016importance}, its inclusion would substantially increase model complexity and computational cost. In this context, our GM transport scheme provides an intermediate complexity alternative balancing realism and efficiency. For exoplanet studies, where the priority lies in capturing first-order climate dynamics rather than detailed regional processes, this simplification remains justified. Nonetheless, future extensions of our model could explore ways to incorporate simplified density-driven processes without the overhead of full ocean dynamics.

The model does not simulate sea ice drift, a simplification made to reduce computation time. On modern Earth, this omission can lead to unrealistically high sea ice accumulation in the Arctic (as seen from our model data in Fig.~\ref{fig:seaice_north_south}), as ice is unable to be transported out of the region and instead thickens and expands in place \citep{codron2012ekman}. While this may be a secondary effect under present-day Earth conditions, sea ice dynamics has been shown to play a critical role in shaping planetary climate regimes. They are essential for simulating transitions between stable climate states on Earth and exoplanets. For example, dynamic sea ice facilitates equatorward transport, promoting Snowball Earth initiation and enhancing the instability of tropical sea ice margins \citep{lewis2007snowball, voigt2012sea}. These processes may be even more important on synchronously rotating planets, where strong day–night contrasts can interact with sea ice transport to produce qualitatively different climate outcomes \citep{yang2020transition}. Future work is needed to assess whether incorporating sea ice drift is feasible within the framework of our model without incurring prohibitive computational costs.

\section{Conclusions}

We introduce a dynamical slab ocean model integrated into a 3-D GCM called the Generic-PCM. The fully parallelised model offers a compromise between the computational efficiency of slab ocean models and the physical realism of fully coupled AOGCMs. Sea ice and snow albedos are parameterised to be spectrally and thickness dependent (Fig.~\ref{fig:albedo_parameterisation}). Unlike imposed-q-flux approaches, our model features emergent ocean heat transport (OHT) including wind-driven Ekman transport, horizontal diffusion, convective adjustment \citep{codron2012ekman, charnay2013exploring}, and a newly implemented Gent–McWilliams (GM) parameterisation for mesoscale eddy mixing (summarised in Fig.~\ref{fig:OHT_Diagram}). Importantly, the realism that comes with OHT-on simulations is at no additional computational cost compared to OHT-off simulations run over the same number of model years.

We first demonstrate the individual (and combined) impacts of our OHT mechanisms (Table~\ref{tab:aquaplanet-experiments}) using a zero obliquity aquaplanet configuration. We find that enabling OHT reduces sea ice coverage and alters tropical SST structure — replacing the unrealistic equatorial SST peak (seen in OHT-off, Fig.~\ref{fig:zonal-temperature-aquaplanet}a) with a relative temperature minimum due to Ekman-induced upwelling (OHT-on, Fig.~\ref{fig:zonal-temperature-aquaplanet}a). This upwelling also suppresses equatorial precipitation, forming a double-banded ITCZ, consistent with theoretical expectations. Figure~\ref{fig:northward-transport-aquaplanet} shows our model aquaplanet's meridional OHT profile, with Ekman transport dominating in the tropics as expected from observations \citep{levitus1987meridional}, and GM and horizontal diffusion peaking near the ice edge — in agreement with previous fully coupled AOGCM studies \citep{marshall2007mean, brunetti2019co}.


We then simulate modern Earth, where our OHT-on configuration produces an annual global average surface temperature of 13°C, close to the NCEP/NCAR reanalysis (14°C), whereas our OHT-off simulation yields a temperature of 12${^\circ}$C. We find that enabling OHT not only drives more realistic SSTs and precipitation patterns (Fig.~\ref{fig:OHT_Tslab1_Prec_Earth}), but also improves seasonal extrapolar SSTs and sea ice extent (Fig.~\ref{fig:SST_SeaIce}). Specifically, the OHT-on case yields extrapolar SSTs within $\approx$ 0.6°C and sea ice coverage within $\approx$ 3 million km² of observations, whereas OHT-off shows much larger biases (upto 2.0°C lower SSTs and 10 million km$^2$ higher annual sea ice coverage than observations). We also report a planetary bond albedo of 0.32, in close agreement with NCEP/NCAR reanalysis (0.31). Our annual means are quite close to observations and we consider this satisfactory by exoplanet modelling standards, given the large uncertainties associated.


Beyond these benchmarks, we find that the GM parameterisation contributes to both horizontal and vertical mixing, with the latter playing a role analogous to convective adjustment (Fig.~\ref{fig:deep-layer-temperature-Dec2025}). Additionally, OHT, driven particularly by the Sverdrup balance, stabilises the overall climate response and reduces hemispheric asymmetries that may arise from initial condition biases (Figs.~\ref{fig:OHT-profiles-comparison}, \ref{fig:hemispheric-asymmetry}, \ref{fig:sverdrup-zonal-map} and \ref{fig:sverdrup-winds}). Given the flexibility of the dynamical slab ocean model, exoplanet-based applications are relatively easy to add. For instance, an additional heat source can be included to simulate geothermal heating, and/or the diffusion coefficients can be adjusted depending on planetary rotation rates. These extensions will be explored in future work.

We simulate well the wind-driven component of oceanic circulation on modern Earth. However, since we do not model the density-driven branch of the MOC, we cannot reproduce the full strength and structure of the observed MOC (Fig~\ref{fig:northward-transport-earth}). Some discrepancies remain between our simulations and observations -- including a southward-shifted ITCZ (Fig.~\ref{fig:OHT_Tslab1_Prec_Earth}), a colder Northern Hemisphere (Fig.~\ref{fig:earth-zon-avg}), and minor deviations in sea ice extent and SSTs (Fig.~\ref{fig:SST_SeaIce}) -- which we analyse and discuss, along with potential strategies for improvement. We also assess the model's sensitivity to the diffusion coefficients and sea ice albedo (Fig.~\ref{fig:earth-zon-avg}), and outline future developments, including the addition of salinity, dynamic sea ice drift, and a varying mixed-layer depth.

The core aim of our model is to strike a balance between physical realism and computational efficiency, which makes it particularly well-suited for large parameter space exploration. By capturing crucial effects of OHT -- without the associated computational cost of solving primitive equations -- the model enables long-timescale simulations and broad sensitivity studies while retaining key processes that shape planetary climates. This work lays the foundation for the continued development of the Generic-PCM, with the goal of enhancing its readiness to better interpret future observations of temperate terrestrial exoplanets from upcoming missions such as RISTRETTO@VLT, PCS@ELT, HWO, and LIFE.


\clearpage

\appendix

\section{\label{sec:OceanPhysicalParameters}Ocean Physical Parameters}

\begin{table}[h!]
\centering
\renewcommand{\arraystretch}{1.2} 
\begin{tabular}{|l|c|}
\hline
\textbf{Parameter} & \textbf{Value} \\
\hline
Freezing seawater temperature, $T_{0}$ [K] & 271.35 \\
\hline
Melting seawater temperature [K] & 273.15 \\
\hline
Mean ice density, $\rho_i$ [kg\,m$^{-3}$] & 917 \\
\hline
Mean snow density [kg\,m$^{-3}$] & 300 \\
\hline
Seawater density [kg\,m$^{-3}$] & 1026 \\
\hline
Ice thermal conductivity, $\lambda$ [W\,m$^{-1}$\,K$^{-1}$] & 2.17 \\
\hline
Snow thermal conductivity, [W\,m$^{-1}$\,K$^{-1}$] & 0.31 \\
\hline
Specific heat capacity of ice (\& snow), $C_i$ [J\,kg$^{-1}$\,K$^{-1}$] & 2067 \\
\hline
Specific heat capacity of seawater, $C_p$ [J\,kg$^{-1}$\,K$^{-1}$] & 3994 \\
\hline
Latent heat of fusion of ice (\& snow), [J\,kg$^{-1}$] & 334000 \\
\hline
Latent heat of sublimation of ice (\& snow), [J\,kg$^{-1}$] & 2834000 \\
\hline
Maximum albedo of ice (\& snow) & 0.65\\
\hline
\end{tabular}
\vspace{1em} 
\caption{Oceanic physical parameters used in the model. Units are given in SI.}
\label{tab:ocean-physical-parameters}
\end{table}

\clearpage

\section{\label{sec:computation-speed}Computation speed}
At the spatiotemporal resolution used in this study (see Sect.~\ref{sec:model-description}), one model year requires approximately 2.75 hours of computation time on 24 cores -- for both, the OHT-on and OHT-off cases. This suggests that Generic-PCM simulations requiring an ocean can enable OHT without incurring additional computational cost for the same number of model years. This efficiency arises from the way OHT is implemented in our dynamical slab ocean model: the diffusion-like processes (i.e., horizontal diffusion and the GM scheme) are relatively inexpensive to compute on coarse grids, and the added overhead per timestep is negligible compared to the cost of the atmospheric model and coupling processes. 

We emphasise that direct quantitative comparisons of wall-clock computation time between climate models must be interpreted with caution, as GCM performance depends strongly on timestep frequency, model heritage, physical parameterisations (particularly radiative transfer), and the computational hardware. Fully coupled AOGCMs such as ROCKE-3D~2.0 \citep{tsigaridis2025rocke} can, in some configurations, achieve shorter per-model-day computation times; for instance, simulation P2SAoM40 in \citet{tsigaridis2025rocke} reports a cost of 0.18 minutes per simulated day, compared to around 0.45 minutes per simulated day for our model. However, differences in timestep frequency and in how often ocean and radiative processes are called make direct one-to-one comparisons difficult. 

Another important practical consideration is not only the wall-clock cost per simulated day, but the number of model years required to reach equilibrium. In our aquaplanet experiments with OHT enabled, equilibrium is typically achieved within 15–40 model years, depending on the initial state. On 24 cores, this corresponds to convergence within 5 days of wall-clock time. Equilibration times for fully coupled AOGCMs are often approximately hundreds to thousands of model years, implying convergence times ranging from weeks to months.

However, at higher horizontal resolutions, OHT-on simulations may become more computationally demanding than OHT-off cases. This is due to the increased cost of additional Laplacian-based calculations, which scale with resolution and can introduce greater parallelisation overhead. But in this study, the ocean physics is called every physical timestep (once every 7.5 or 15 model minutes, see Sect.~\ref{sec:model-description}). This frequency could be reduced -- for example, to a few calls per model day or even just once daily -- with minimal impact on physical accuracy and a notable gain in computational efficiency. Nonetheless, even at high spatial resolutions, we expect our dynamical slab ocean model to remain significantly faster than a fully coupled AOGCM at equivalent resolution since it avoids solving the full set of fluid prognostic equations, which are typically the main computational bottleneck.

\clearpage

\section{\label{sec:annually-averaged-maps}Annually averaged maps}
\begin{figure}[h]
    \centering
    \begin{subfigure}[b]{0.8\textwidth}
        \centering
        \includegraphics[width=\textwidth]{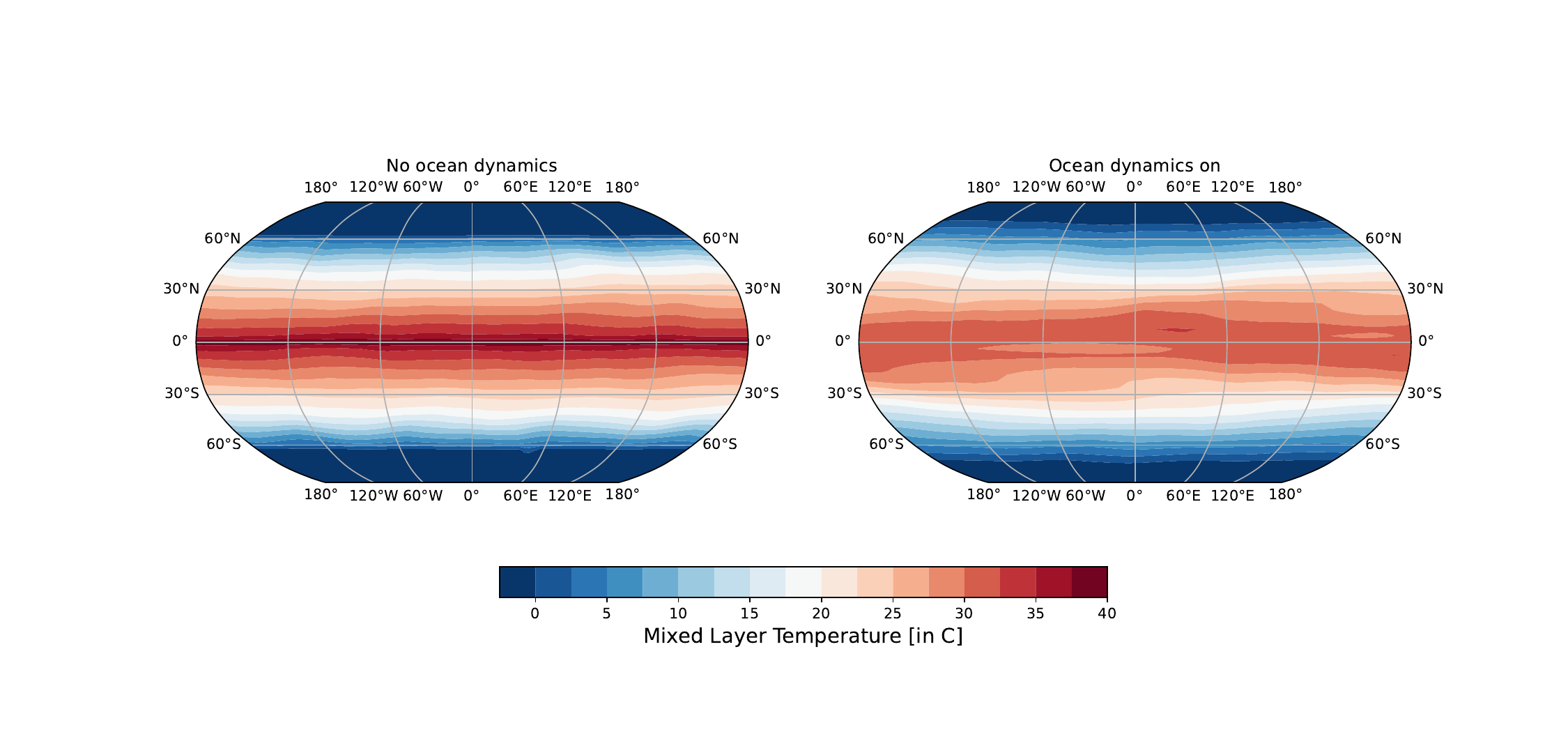}
    \end{subfigure}

    \vspace{-4em} 

    \begin{subfigure}[b]{0.8\textwidth}
        \centering
        \includegraphics[width=11cm]{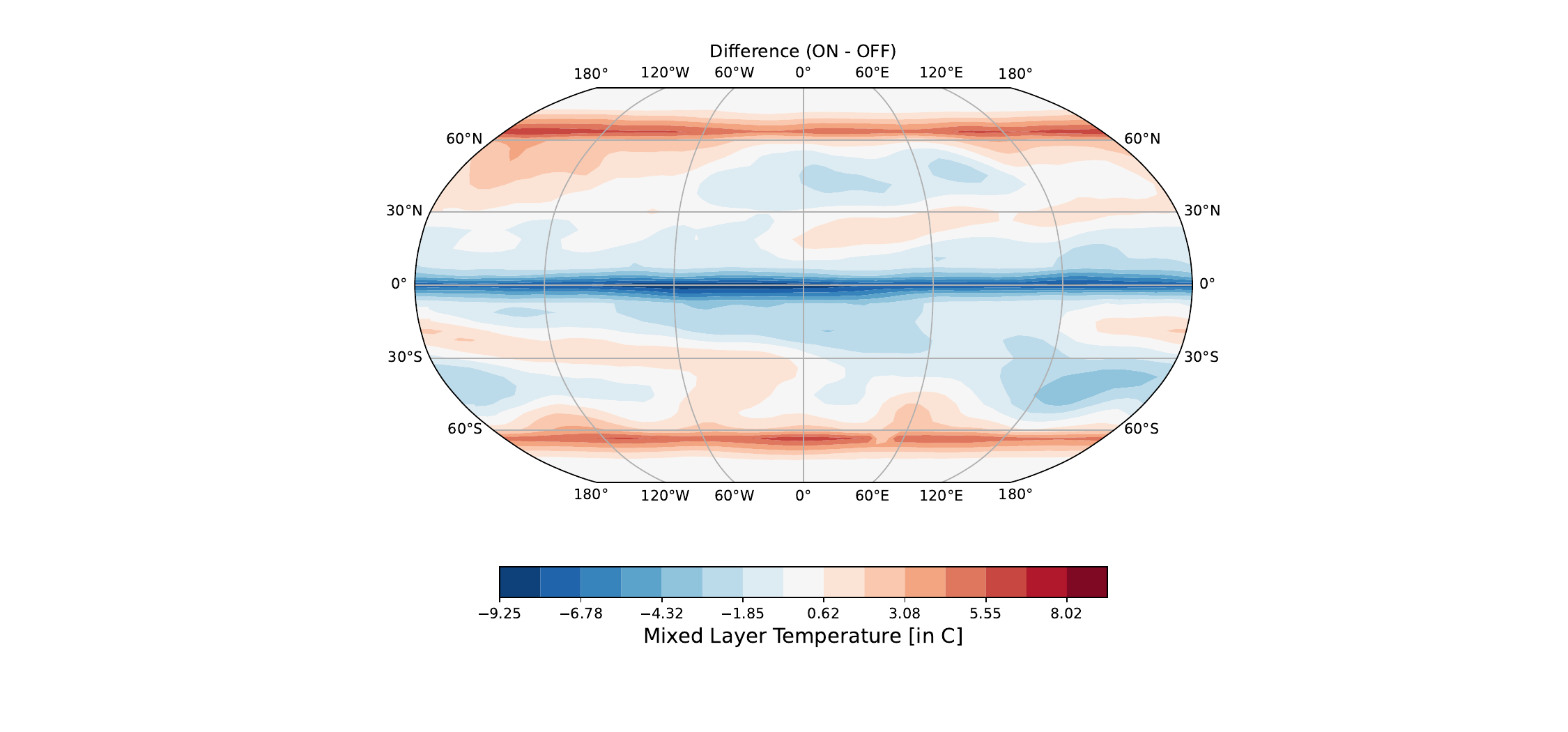}
    \end{subfigure}
    
    \vspace{-2em} 
    
    \caption{\label{fig:mixed-layer-temperature-aquaplanet}Annually averaged sea surface temperature (SST) maps for the zero-obliquity aquaplanet simulations. The top-left panel shows SST for Case 1 (no ocean heat transport, OHT), while the top-right panel shows SST for Case 7 (all OHT enabled). The bottom panel displays the difference between the two (Case 7 - Case 1), highlighting the impact of OHT. Notable features include equatorial upwelling of cold water and enhanced warming at higher latitudes.}
    \label{fig:combined-plot}
\end{figure}

\clearpage

\section{\label{sec:atmospheric-profiles}Atmospheric profiles}
\begin{figure}[h]
\centering
\includegraphics[width=15cm]{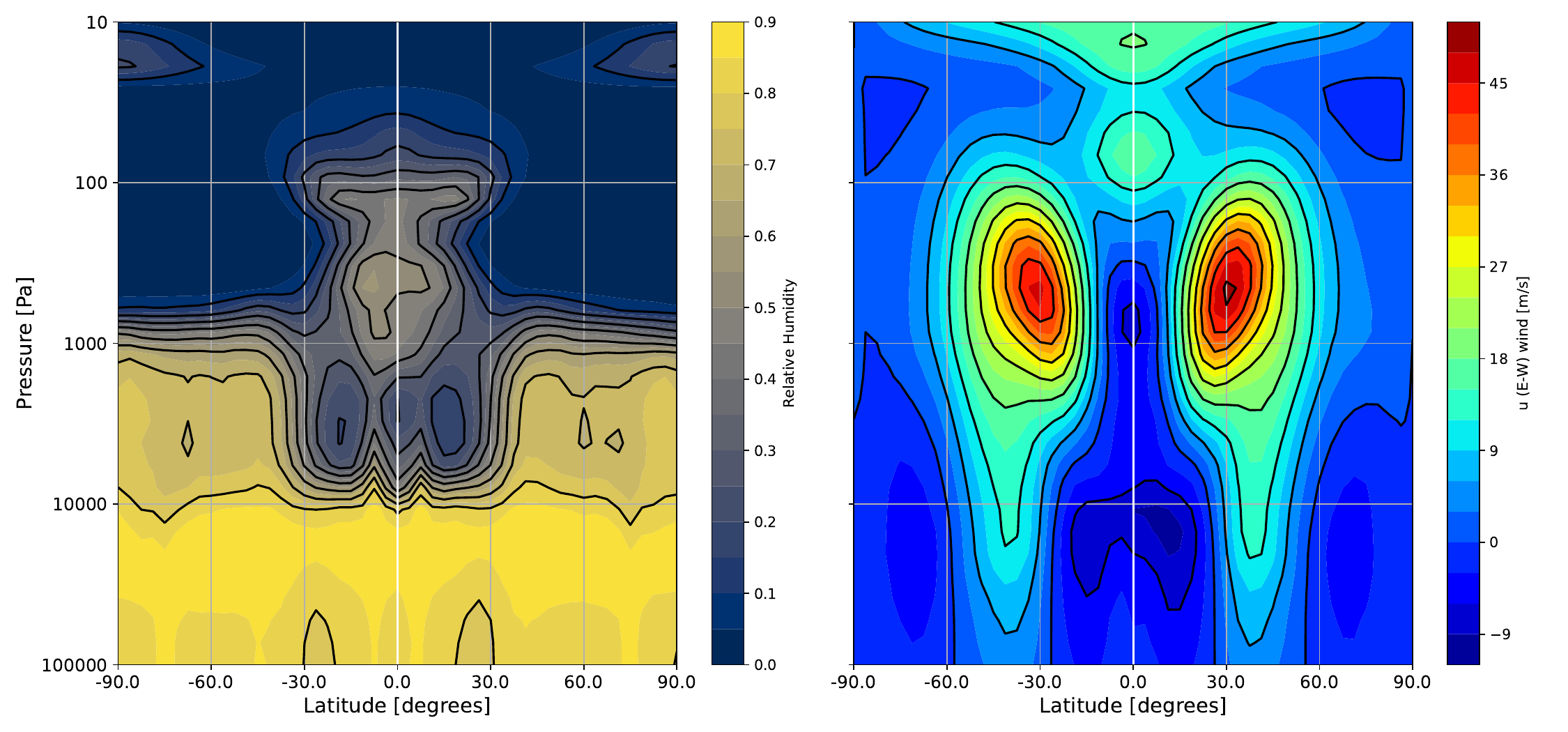}
\caption{\label{fig:aquaplanet-atmosphere-profiles}Annually averaged atmospheric profiles for the zero-obliquity aquaplanet simulations. The left plot shows the relative humidity of the atmosphere as a function of pressure levels and latitude for the Case 7 simulation (all ocean heat transport on). The plot on the right shows the same simulation's zonal winds, with the mid-latitude jet streams clearly visible.}
\end{figure}

\clearpage

\section{\label{sec:appendix-seasonal-evolution}Seasonal evolution of modern Earth's sea ice coverage}
\begin{figure}[h]
\centering
\includegraphics[width=15cm]{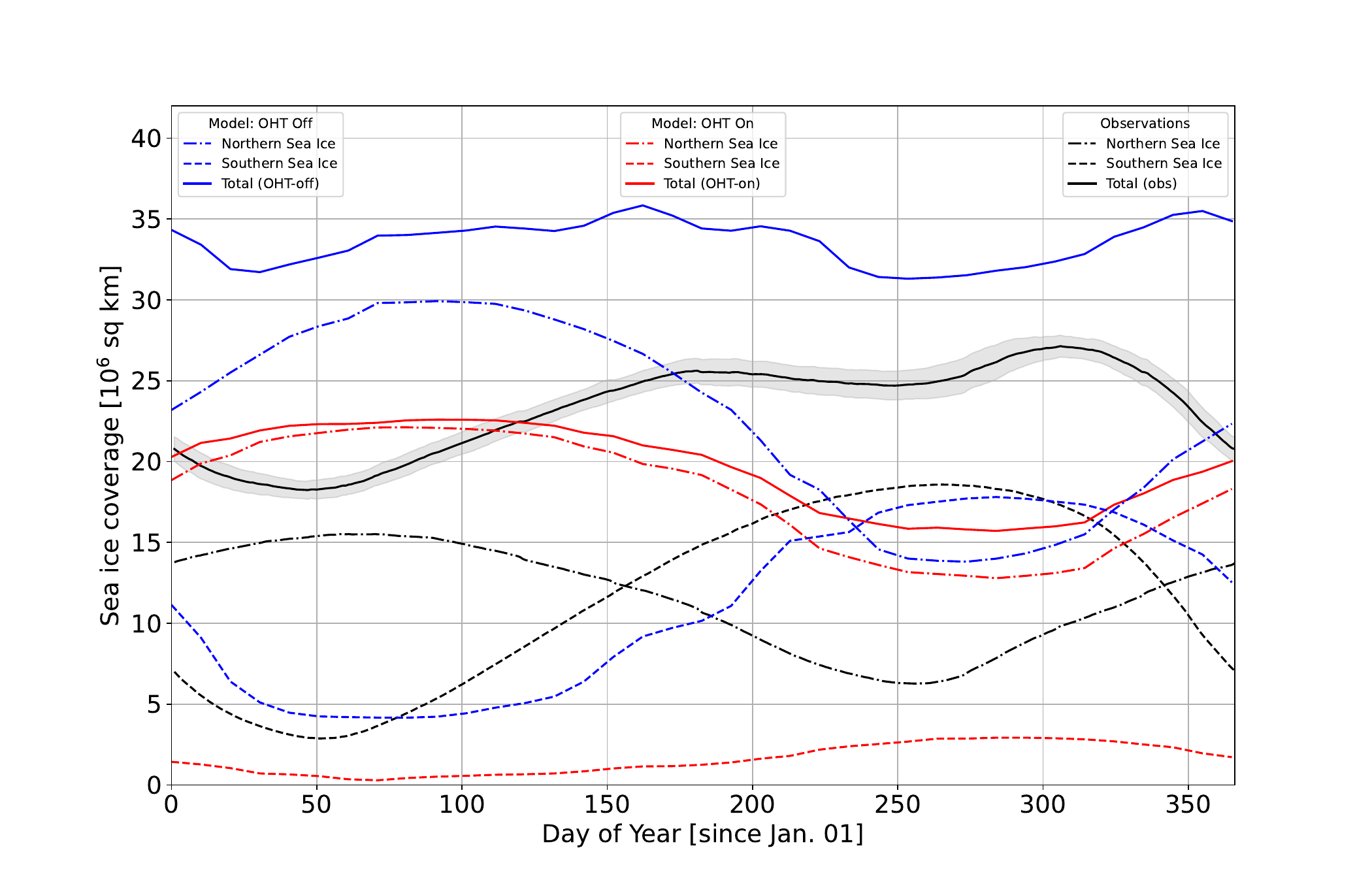}
\vspace{-1em} 
\caption{\label{fig:seaice_north_south}Seasonal evolution of global sea ice coverage for NSIDC observations (black), the model with ocean heat transport enabled (red), and disabled (blue). For each case, the dashed lines represent Southern sea ice and the dash-dotted lines the Northern sea ice.}
\end{figure}

\clearpage

\section{\label{sec:appendix-sverdrup-balance}Surface wind response to the Sverdrup balance scheme}
\begin{figure}[h]
\centering
\includegraphics[width=15cm]{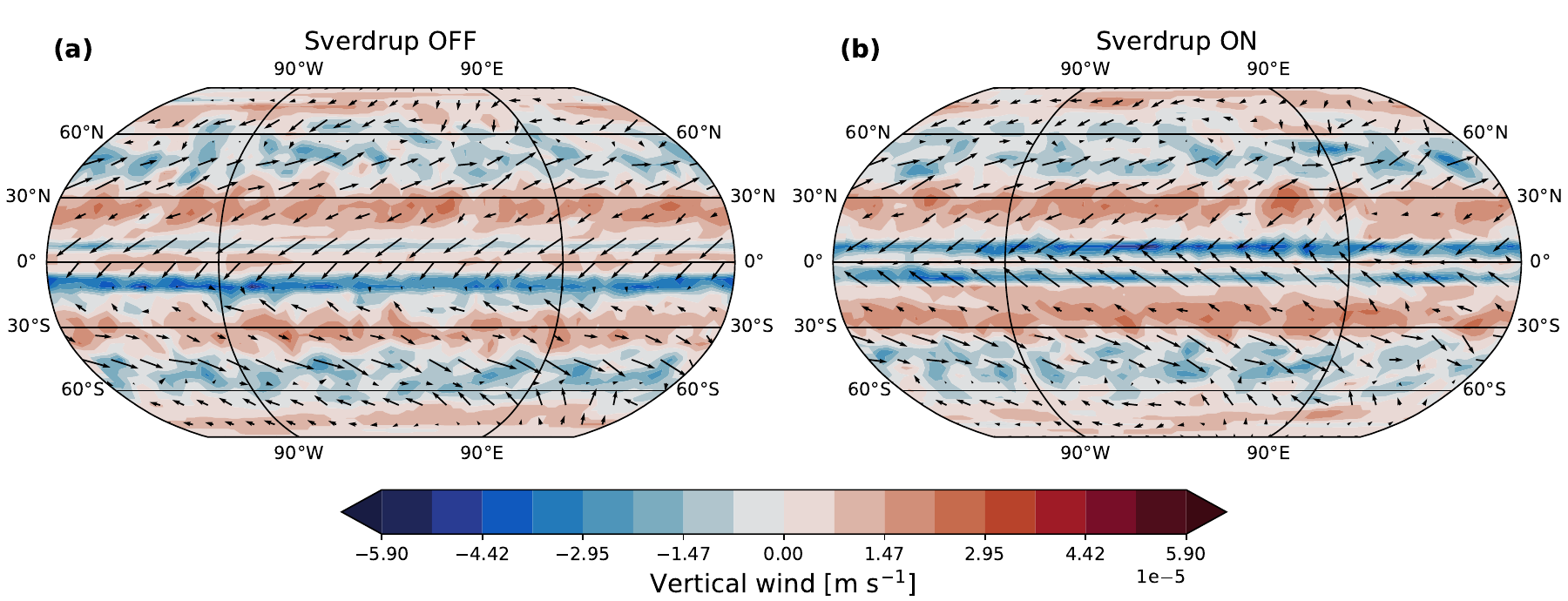}
\vspace{-1em} 
\caption{\label{fig:sverdrup-winds}Decadal-mean net surface wind vectors for simulations \textbf{(a)} without and \textbf{(b)} with Sverdrup balance. For the latter, near-equatorial transport responds to the wind stress curl rather than the frictional meridional component, producing a more symmetric wind field than for the case without the Sverdrup balance.}
\end{figure}

\clearpage

\section{\label{sec:appendix-comparison-with-codron}Comparison with \citet{codron2012ekman}}

\subsection{\label{subsec:direct-overlays-codron-sst-prec}Direct overlays of SST and precipitation profiles}
\begin{figure}[h]
\centering
\includegraphics[width=15cm]{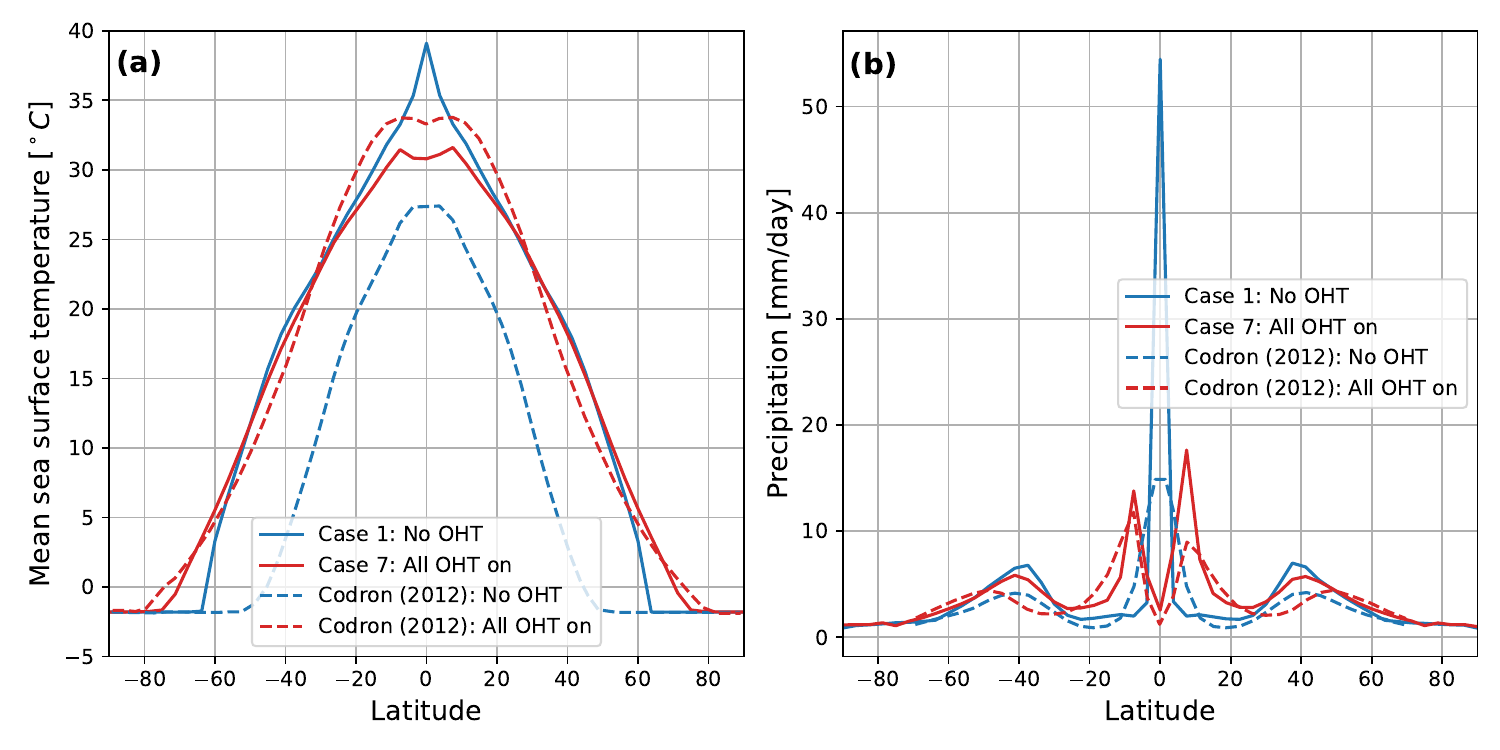}
\vspace{-1em} 
\caption{\label{fig:aquaplanet-temperature-precipitation-compare-codron}Decadal-mean zonally averaged plots for \textbf{(a)} sea surface temperature (SST) for OHT-off (blue) and OHT-on (red) configurations in the present study (solid lines) and \citet{codron2012ekman} (dashed lines), and
(b) precipitation. Differences between the two implementations reflect changes in the atmospheric model and resolution between the studies.}
\end{figure}

Interpreting the relative positions of features in the SST and precipitation profiles between \citet{codron2012ekman} and the present study, and attributing them to specific physical mechanisms, is challenging due to substantial evolution in the GCM (including its atmospheric model) between the two implementations. This is already evident when comparing the OHT-off configurations, which isolate the atmospheric component: the zonal-mean SST profiles in Fig.~\ref{fig:aquaplanet-temperature-precipitation-compare-codron}a (solid and dashed blue lines in the present work and \citealt{codron2012ekman}, respectively) differ markedly between the two studies.

The large difference in equatorial temperatures in the OHT-off cases (Fig.~\ref{fig:aquaplanet-temperature-precipitation-compare-codron}a) directly affects the magnitude of the equatorial precipitation maxima (Fig.~\ref{fig:aquaplanet-temperature-precipitation-compare-codron}b). In \citet{codron2012ekman}, enabling OHT leads to a slight poleward shift of the storm tracks relative to the OHT-off case. A similar modest shift is observed in our study. However, it is unclear whether this reflects a physical response or a resolution-related artefact.

More generally, differences in profile shapes between \citet{codron2012ekman} and this study likely reflect differences in atmospheric physics, particularly cloud / precipitation parameterisations, as well as the resolution (leading to stronger / weaker baroclinic eddies). These combined effects complicate direct one-to-one comparisons of individual climatic features.

\subsection{\label{subsec:vertical-ocean-structure}Vertical ocean structure}
It is important to assess whether the improvements obtained with the present model configuration reflect the inclusion of GM transport, or whether similar behaviour could be achieved with an enhanced horizontal diffusion coefficient (25000~m$^2$~s$^{-1}$), Ekman transport and convective adjustment, like in \citet{codron2012ekman}. We compare this latter configuration with the recommended ocean setup of the present model (Case 7b), which employs a smaller horizontal diffusion coefficient (8000~m$^2$~s$^{-1}$), GM transport (2000~m$^2$~s$^{-1}$), Ekman transport and convective adjustment.

\begin{figure}[h]
\centering
\includegraphics[width=11cm]{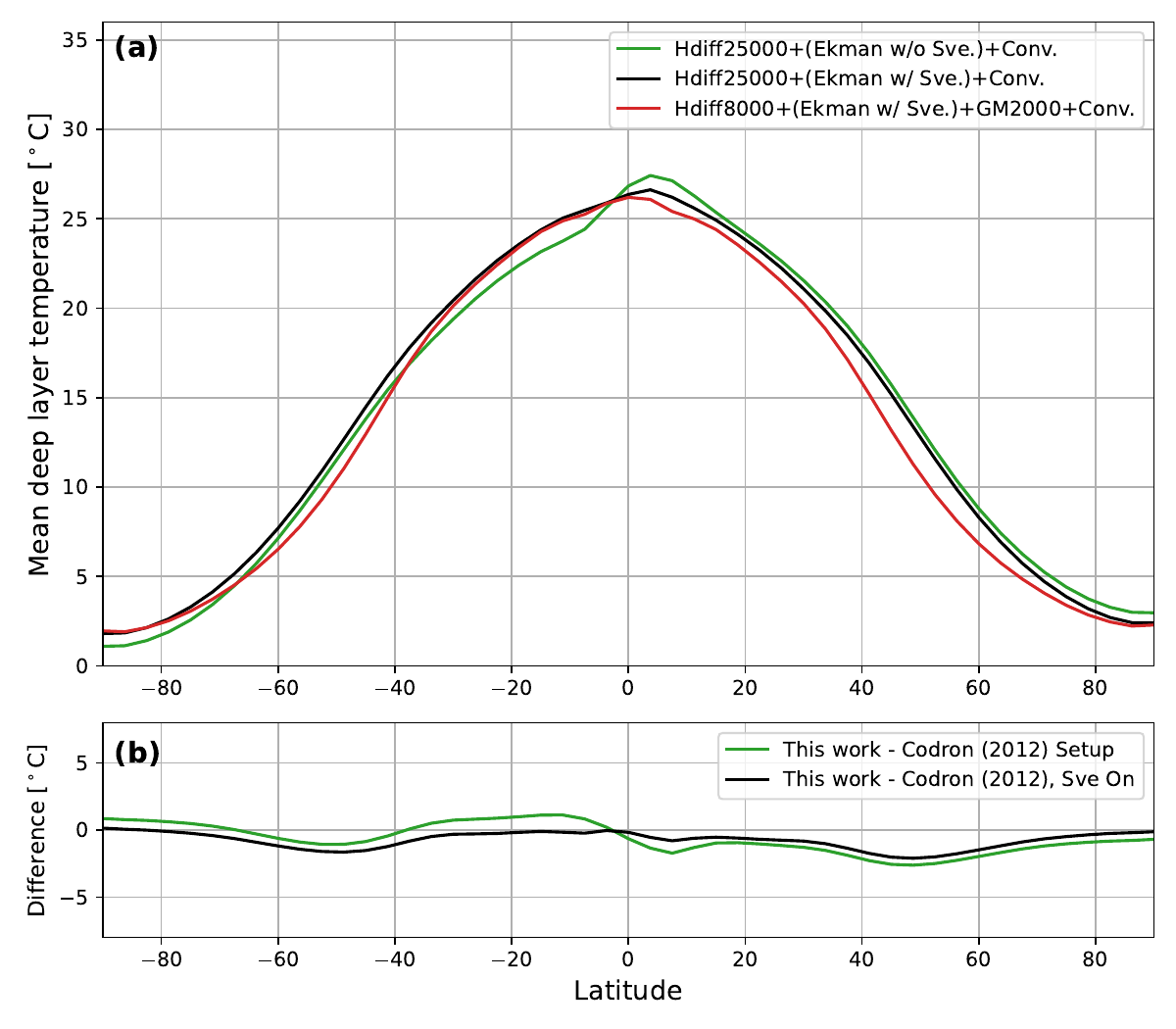}
\vspace{-1em} 
\caption{\label{fig:deep-layer-temperature-comparison-Codron}Influence of Gent–McWilliams (GM) transport on deep-ocean thermal structure: \textbf{(a)} Decadal-mean zonally averaged deep layer temperature for the recommended configuration of the present model (red; including GM) and Codron-style configurations (green: original setup; black: Codron setup with Sverdrup balance enabled).
\textbf{(b)} Temperature differences relative to the GM configuration.}
\end{figure}

Figure~\ref{fig:deep-layer-temperature-comparison-Codron}a shows the zonally averaged deep layer temperature profiles for the tested configurations. Comparison with the \citet{codron2012ekman} configuration (green line) is complicated by the pronounced hemispheric asymmetry that now develops in the absence of the Sverdrup balance, as explained in Sect.~\ref{subsec:hemisphere-asymmetry}. To isolate the effects of the parameter choices, we perform a simulation using the Codron-like setup but with the Sverdrup balance enabled (black line). We see that both these Codron-style configuration(s) produce(s) systematically warmer deep layer temperatures at mid-latitudes than the present configuration with GM (red line). The difference of the \citet{codron2012ekman} configurations with the current recommended setup is shown in Figure~\ref{fig:deep-layer-temperature-comparison-Codron}b. Here, we see that the simulation with GM transport is approximately 2$^\circ$C cooler in the mid-latitudes (black line). This comparison again demonstrates that neither enhanced horizontal diffusion nor convective adjustment alone can fully reproduce the thermal structure obtained when GM is included. It further indicates that by modifying large-scale stratification, GM transport exerts a distinct influence on the climate.

\clearpage












{\footnotesize \textit{Code and data availability.} All model simulations presented in this study were performed using revision 3423 of the Generic-PCM \citep[][latest revision available at \url{http://svn.lmd.jussieu.fr/Planeto/trunk/LMDZ.GENERIC/}]{generic-pcm-svn}. A frozen version of the model code, together with the initial conditions and the GCM outputs from the final model year of the simulations conducted in the manuscript, is archived on Zenodo \citep[][\url{https://doi.org/10.5281/zenodo.18771594}]{bhatnagar2025zenodo}. The concept DOI (\url{https://doi.org/10.5281/zenodo.16417154}) represents all versions of the repository. Additional data supporting this article can be provided upon request to the corresponding author.
\\The primary routines related to the dynamical slab ocean model -- \texttt{ocean\_slab\_mod.F90}, \texttt{slab\_heat\_transp\_mod.F90},\\ \texttt{physiq\_mod.F90} and \texttt{hydrol.F90} -- are located within the physics directory of the Generic-PCM source code. Model documentation is provided on the Generic-PCM Wiki page \citep[][\url{https://lmdz-forge.lmd.jussieu.fr/mediawiki/Planets/index.php/Overview_of_the_Generic_PCM}]{generic-pcm-overview} and practical guidance on the dynamical slab ocean model is available on its dedicated page \citep[][\url{https://lmdz-forge.lmd.jussieu.fr/mediawiki/Planets/index.php/Slab_ocean_model}]{slab_ocean_documentation}. 
\\All reanalysis and observational datasets used in this study are publicly available. NCEP/NCAR Reanalysis data \citep{kalnay1996} and NOAA DOISST V2.1 High Resolution Dataset \citep{huang2021improvements}  are provided by the NOAA Physical Sciences Laboratory, Boulder, Colorado, USA, and can be accessed at \url{https://psl.noaa.gov/data/gridded/data.ncep.reanalysis.html} and \url{https://psl.noaa.gov/data/gridded/data.noaa.oisst.v2.highres.html}, respectively. ECMWF ERA-15 reanalysis data \citep{gibson1997era15} are available through the Copernicus Climate Data Store at \url{https://cds.climate.copernicus.eu}. Observational sea ice extent data were obtained from the NSIDC Sea Ice Index, Version 3 \citep[][\url{https://doi.org/10.7265/N5K072F8}]{fetterer-nsidc}.} 

\vspace{30px}

{\footnotesize \textit{Video supplement.} The video file available on Zenodo \citep[][\url{https://doi.org/10.5281/zenodo.18771594}]{bhatnagar2025zenodo} illustrates the temporal evolution of SST for the zero-obliquity aquaplanet simulations.} 

\vspace{30px}


{\footnotesize \textit{Author contributions.}
S.B., E.B., M.B., J.K., and M.T. developed the study's conceptualisation. S.B., F.C., E.M., and M.T. contributed to the model code. S.B., E.B., M.B., and J.K. developed the methodology and conducted the analysis. S.B. performed the simulations, related investigation and wrote the original draft. All authors contributed to the review and editing of the manuscript. E.B., M.B., and J.K. supervised the project and related administration. E.B. and J.K. acquired the project funding.
}
\vspace{30px}

{\footnotesize \textit{Competing interests.}
The authors declare that they have no conflict of interest.
} 

\vspace{30px}

{\footnotesize \textit{Acknowledgements.}
This work has been carried out within the framework of the NCCR PlanetS supported by the Swiss National Science Foundation under grants 51NF40\_182901 and 51NF40\_205606. 
E.B. acknowledges the financial support of the SNSF (grant number: 200021\_197176 and 200020\_215760). 
M.B. acknowledges the financial support from the Swiss National Science Foundation (Sinergia Project
No. CRSII5\_213539). 
M.T. acknowledges support from the Tremplin 2022 program of the Faculty of Science and Engineering of Sorbonne University, and the High-Performance Computing (HPC) resources of Centre Informatique National de l'Enseignement Supérieur (CINES) under the allocations No. A0140110391 and A0160110391 made by Grand Équipement National de Calcul Intensif (GENCI).
G.C. acknowledges the financial support of the SNSF (grant number: P500PT\_217840).
The authors thank the referees for their constructive comments and suggestions, which led to additional analyses and have improved the manuscript.
We also thank the Generic-PCM team for the teamwork development and improvement of the model.
S.B. thanks the GAP-Nonlinearity and Climate team, the Dynaclim team and Stéphane Goyette for helpful discussions. 
The computations were performed at the University of Geneva's HPC clusters of Baobab and Yggdrasil. This research has made use of NASA’s Astrophysics Data System. We confirm that all figures included in this manuscript are original and created by the authors.
}







\bibliographystyle{copernicus}
\bibliography{paper-updated13Apr.bib}

\end{document}